\documentclass[aps,pra,superscriptaddress]{revtex4}
\usepackage[intlimits]{amsmath}
\usepackage{amsfonts}
\usepackage{psfrag}
\usepackage{subfigure}
\usepackage{enumitem}
\usepackage[usenames]{color}
\advance\voffset by -1cm
\advance\hoffset by -0.5cm
\newcommand\be            {\begin{equation}}
\newcommand\bea           {\begin{equation}\begin{array}l\displaystyle}
\newcommand\ee            {\end{equation}}
\newcommand\bes           {\begin{subequations}}
\newcommand\esu           {\end{subequations}}

\newcommand{\bigx}[1]{\bBigg@{#1}}

\def\3pt#1#2#3{{\langle{#1}\vert{#2}\vert{#3}\rangle}}

\newcommand\doi[2]        {\href{http://dx.doi.org/#1}{#2}}

\newcommand{\EQ}{\begin{equation}}
\newcommand{\EN}{\end{equation}}
\usepackage{epsfig}
\usepackage{color}
\usepackage{psfrag}
\usepackage{amsmath}
\usepackage{graphicx}
\usepackage{amsfonts}
\usepackage{amssymb}
\begin{document}
\bibliographystyle{plainnat}

\title{{\Large {\bf Bound States of Majorana Fermions in \\ Semi-classical Approximation}}}

\author{G. Mussardo}
\affiliation{SISSA and INFN, Sezione di Trieste, via Bonomea 265, I-34136, 
Trieste, Italy}
\affiliation{International Centre for Theoretical Physics (ICTP), 
I-34151, Trieste, Italy}
\affiliation{International Institute of Physics, Natal, Brasil}

\begin{abstract}
\noindent
We derive a semi-classical formula for computing the spectrum of bound states made of Majorana fermions in a generic non-integrable 2d quantum field theory with a set of degenerate vacua. We illustrate the application of the formula in a series of cases, including an asymmetric well potential where the spectra of bosons and fermions may have some curious features. We also discuss the merging of fermionic and bosonic spectra in 
the presence of supersymmetry. Finally, we use the semi-classical formula to analyse the evolution of the particle spectra in a class of 
non-integrable supersymmetry models.    

\vspace{3mm}
\noindent
Pacs numbers: 11.10.St, 11.15.Kc, 11.30.Pb

\end{abstract}
\maketitle

\section{Introduction}
\label{sec:intro}
\noindent
The experimental advances in neutron scattering and material design, as well as in the manipulation of ultra-cold atom systems, 
have recently permitted to realise and study in detail the class of universality of some of the most important low-dimensional quantum field theories. This is the case of the 2d Ising model in a magnetic field \citep{Zam,DMIsing}, a model not only important for its 
physical and historical relevance but also for its deep relation with beautiful mathematical object such as $E_8$ Lie Algebra. 
The class of universality of this model has been experimentally accomplished by designing a quantum spin chain made of Co${\rm Nb}_2 {\rm O}_6$ (cobalt niobate) whose properties were subsequently analysed by neutron scattering \citep{Coldea}. The experimental analysis has confirmed, in particular, the theoretical prediction about the existence of three bound states below threshold \citep{Zam}, with their weights in the spectral density of spin-spin correlation function being in perfect in agreement with their theoretical determination \citep{DMIsing}. Another example of class of universality is the celebrated Sine-Gordon model \citep{zamzam,ffstructure1,ffstructure2}, recently experimentally realised  in terms of a cold-atom set-up and then studied through interference patterns \citep{Schweigler}. This experimental work has permitted, in particular, to study the phase correlations, to characterise the topologically distinct vacua and the interactions of the excitations above them. 

Given the present trend of such a theoretical and experimental advance, one may hope to witness progress in the next future on one 
of the long-standing aspects of quantum field theories (QFT), alias how to determine the spectrum of their physical excitations. This is a genuine dynamical problem, since it is a well known that the QFT Lagrangians may employ fields whose particle excitations 
are actually not present in the physical spectrum. 

The most notorious case is probably Quantum Chromodynamics, where the quarks and gluons of its Lagrangian are absent from the spectrum which is given, instead, only by their hadronic and mesonic bound states (see, for instance, \citep{Halzen}). Let's also remind that for theories as QCD the only reliable way known up to now to get the masses of the hadron/meson particles is through numerical approaches -- such as Montecarlo simulations of lattice gauge theories (see e.g. \citep{Montecarlo}).  

Another famous example -- which has the advantage to be exactly solvable, therefore it neatly clarifies what kind of problem is to deal  
with the spectrum of a QFT -- is the aforementioned Sine-Gordon model: it describes the dynamics of a two-dimensional bosonic field $\varphi(x)$ with interaction given by the potential 
\EQ
U_{SG}(\varphi)\,=\,\frac{m_0^2}{\beta^2} \left[1 - \cos(\beta \varphi)\right] \,\,\,,
\label{SG}
\EN 
where $m_0$ is a mass-like parameter and $\beta$ is a coupling constant. Such a theory has an infinite 
number of degenerate vacua $\mid n\,\rangle$ ($n = 0,\pm 1,\pm 2,\ldots$), localised at $\varphi_n^{(0)} 
\,=\,2 \pi \,n/\beta$, each of them with the same curvature of the potential $U_{SG}(\varphi)$ in these 
points, namely $\omega^2 = m_0^2$.  In terms of the usual and familiar perturbative arguments, 
one would be then tempted to conclude that each vacuum has always, at least, one neutral excitation above it. 
However this conclusion may be false. Indeed the exact $S$-matrix of this model \citep{zamzam} shows that the 
situation may be rather different and even quite surprising: indeed, if $\beta^2 > 4 \pi$ such a particle 
-- the very same particle described by the field $\varphi(x)$ itself -- does not exist! When this happens, as in the analogous case of QCD, 
one shall conclude that the spectrum of the theory is not made of the field/particle entering the Lagrangian but consists of other 
excitations that have to be properly identified. 

We brought these two examples simply to argue that the determination of the spectrum of an interactive QFT may be    
quite a difficult and non-perturbative problem which, apart from numerical methods, can be in general addressed  
by few analytic approaches as, for instance, the Bethe-Salpeter equation \citep{Bethe-Salpeter}. There are cases, however, 
where life is easier. This happens, for instance, for two-dimensional integrable QFT's. In this case, indeed, the infinite 
number of conservation laws implies the elasticity and the factorization of their $S$-matrix \citep{Zam,zamzam}, 
which can be then exactly determined in innumerable cases (see, for instance, \citep{GMbook}); and, of course, 
when the exact expression of the $S$-matrix is known, the spectrum of the corresponding theory tidely derives 
from a proper identification of its poles. 
 
Luckily enough, certain progress can be also made when the two-dimensional theories are not integrable, depending of course on the 
degree of \textquotedblleft {\em non-integrability}\textquotedblright, so to speak. For instance, if the theory in question may be regarded as small deformation of an integrable model, one can then apply the Form Factor Perturbation Theory (FFPT) \citep{DMS,DM,CM1,decay,GMSUSYKINK} 
to follow the evolution of the spectrum moving away from the integrable direction. Eventually combining this method together with the 
efficient numerical method of Truncated Conformal Space Approach (TCSA) \citep{YZ}, a big deal of information can be obtained on a series of 
aspects of QFT, such as confinement of topological excitations, adiabatic shift of the mass of non-topological excitations, decays of higher mass particles, presence of resonances etc. (see, for instance, \citep{DMS,DM,CM1,decay,GMSUSYKINK,CM2,DG,ZamSpect,Rut,MusTak,Takacsall}). It is also worth mentioning that for gauge theories in two dimensions, the Bethe-Salpeter formalism finds an elegant application in the computation of the mass spectrum of the mesons \citep{Hooft}. 

In addition to all methods above mentioned (FFPT, TCSA and Bethe-Salpeter equation), there exists another one -- the so-called {\em semi-classical method} \citep{DHN} -- which has the advantage to rely neither on integrability nor of small breaking thereof. While, in general, the application of this method may be quite intricated, it drastically simplifies if applied to the analysis of QFT's with vacua degeneracy and topological kink excitations, as shown in \citep{GJ,JR,FFvolume,GMneutral}. In particular, properly refining and interpreting the relevant matrix elements on kink states originally stated in the paper by Goldstone and Jackiw \citep{GJ}, in the paper \citep{GMneutral} we have shown that it is possible to carry out a thorough study of the semi-classical spectrum of neutral bound states in purely bosonic kink-like theories. 

The aim of this paper is thus to extend the analysis of the paper \citep{GMneutral} to the case in which there are also Majorana fermions, taking as our 
starting point the paper by Jackiw and Rebbi \citep{JR}. As we will see, there are a series of quite interesting situations that can analysed with such a formalism, such as: (a) QFT with the bosonic part of the potential having symmetric or asymmetrical potential wells; (b) QFT with kink excitations which are invariant under Supersymmetry; (c) QFT which are integrable or non-integrable.  

To make the analysis of this paper self-contained, in Section \ref{sec:purelybosonic} and \ref{sec:semiclassboson} we will briefly review the main results relative to the semi-classical treatment of bosonic theories. In Section \ref{sec:semiclassfermion} we address the topological excitations of QFT with fermions and the new vacua structure induced by them. Section \ref{fermionsemiclfor} is the central part of the paper, where one can find our 
semi-classical formula for computing the spectrum of fermions in theories with degenerate ground states. The remaining sections 
are devoted to the analysis of a series of relevant examples, among which: (i)  QFT with symmetric and asymmetric vacua (Sections \ref{symmetriccase} and \ref{asym}) (ii) Supersymmetric theories (Sections \ref{SUSY} and \ref{generalSUSY}); (iii) Integrable  (Section \ref{integrableSUSY}) and non-integrable Supersymmetric theories (Sections \ref{nonintegrableSUSY} and \ref{vacuaSUSY}). Our conclusions are then discussed in Section \ref{conclusions}.

\section{Purely bosonic theories}
\label{sec:purelybosonic}
\noindent
Let's initially consider a purely two-dimensional bosonic theory since it will set the stage for all our future considerations. 
Such a theory concerns with a scalar real field $\varphi(x)$ and a Lagrangian density given by 
\EQ
{\cal L} \,=\,\frac{1}{2} (\partial_{\mu} \varphi)^2 - U(\varphi,\{\lambda\}) \,\,\,,  
\label{Lagrangian}
\EN 
where $\lambda$ is some coupling constant. The case of our interest is when the potential $U(\varphi,\{\lambda\})$ possesses several degenerate minima, localises at $\varphi_a^{(0)}$ ($a =1,2,\ldots,n$), with $U(\varphi^{(0)}_n) = 0$ as shown in Figure 1. In QFT language, these minima identify different vacua $ \mid {\bf a}\,\rangle$ and therefore the basic excitations of such a theory are the topological configurations 
that interpolate between any neighbouring pair of them, namely kinks and anti-kinks. Let's see in more details what they consist of.
\begin{figure}[t]
\vspace{10mm}
\psfig{figure=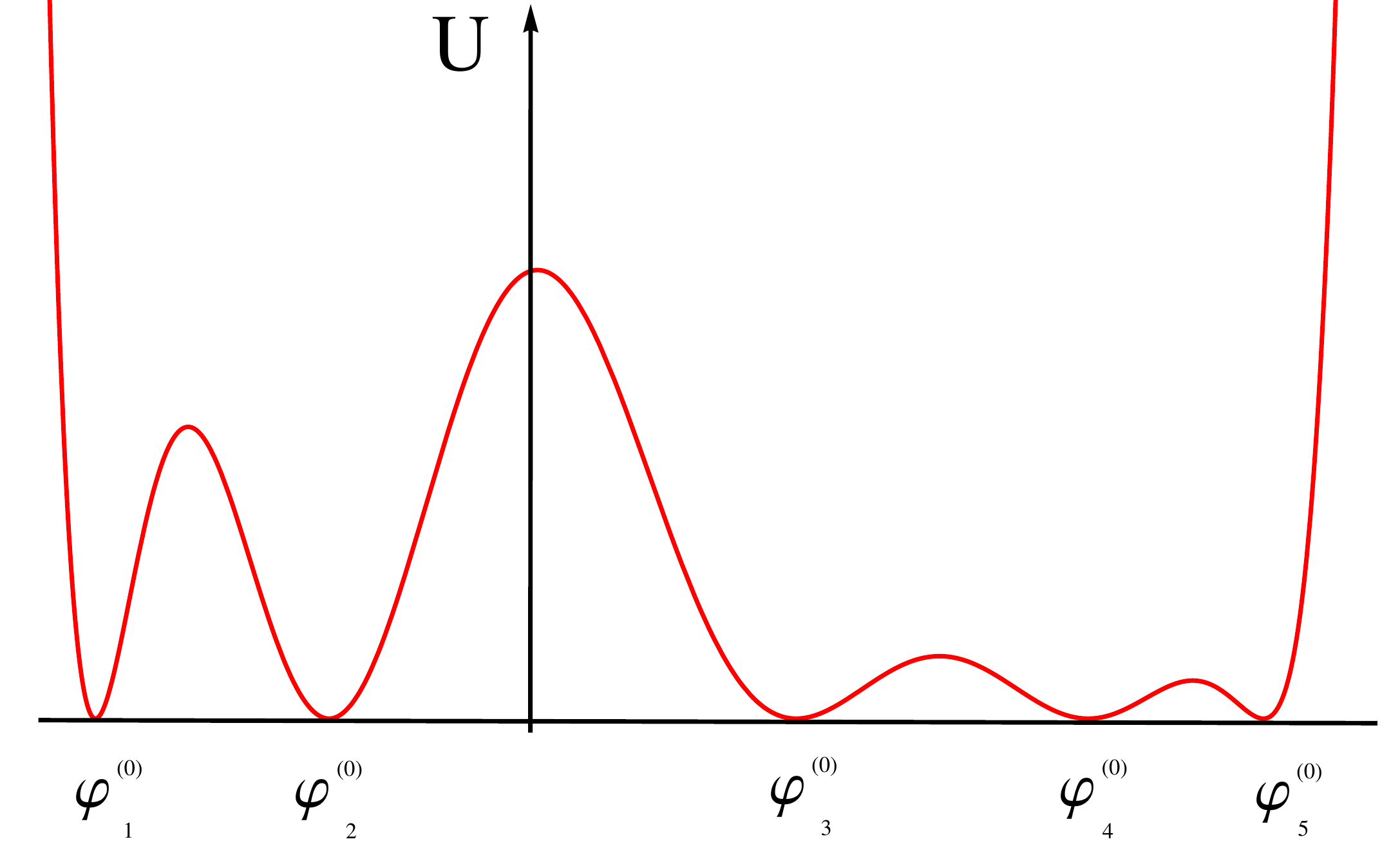,height=4cm,width=6cm}
\caption{{\em Potential $U(\varphi)$ of a bosonic quantum field theory with kink excitations.}}
\label{potential}
\end{figure}
Semi-classically such excitations are identified as the static solutions of the equation of motion, alias
\EQ
\partial^2_x \,\varphi(x) \,=\,U'[\varphi,\{\lambda\}] \,\,\,, 
\label{static}
\EN
with boundary conditions 
\EQ
\left\{
\begin{array}{c}
\varphi(-\infty) = \varphi_a^{(0)} \\
\varphi(+\infty)= \varphi_{b}^{(0)}
\end{array} \right.
\hspace{5mm}\,,\hspace{5mm}  
b = a \pm 1 \,\,\,.
\EN
Let's $\varphi_{ab}(x)$ be the classical solutions which interpolate between $\varphi^{(0)}_a$ (at $ x = - \infty$) and $\varphi^{(0)}_b$ (at $x = + \infty$): with $b = a +1$,  $\varphi_{ab}(x)$ is conventionally called {\em kink} while $\varphi_{ba}(x)$ {\em anti-kink}. In the following, however, we will often 
find easier to use the word \textquotedblleft kink \textquotedblright to denote both of them, especially in discussions of general context. It is worth stressing that these kink configurations can be equivalently obtained as solutions of the simpler first order differential equation (with the same boundary conditions as above) 
\EQ
\frac{d\varphi}{d x} \,=\,\pm \sqrt{2 U(\varphi)} \,\,\,, 
\label{kinkequation}
\EN 
where the $\pm$ signs refer to the kink and the anti-kink respectively.  

\vspace{3mm}
\noindent
{\em {\bf Basic properties of the kink/anti-kink solutions}}. It is useful to remind some features of the kink/anti-kink solutions particularly relevant for the physical consequences they induce. In the following, unless differently stated, $b = a+1$. 

\begin{enumerate}
\item As it will become soon evident, it is important to know how $\varphi_{ab}(x)$ approaches the two asymptotic values $\varphi^{(0)}_{i}$ ($i = a, b$). This problem was solved in \citep{GMneutral}, where it was shown that when the curvature $\omega_i\equiv \sqrt{U''(\varphi^{(0)}_i)}$ ($ i = a,b$) at the two vacua is different from zero, these  approaches happen in an exponential way. The simplest way to show this result is to introduce the variables $\eta_{i} \equiv \left(\varphi(x) - \varphi^{(0)}_{i}\right)$ ($ i = a,b$) and expand the right-hand side of eq.\,(\ref{kinkequation}) (for our choice of the indices $a$ and $b$, the one with positive sign) in power of $\eta_i$ in the vicinity of the minimum $\varphi^{(0)}_a$ (for $x \rightarrow - \infty$) or $\varphi^{(0)}_b$ (for $x \rightarrow + \infty$) so that 
\EQ
\frac{d\eta_i}{d x} \,=\,\omega_i\, \eta_i + \alpha^{(i)}_2 \eta_i^2 + \alpha^{(i)}_3 \eta_i^3 + \ldots \,\,\,,  
\label{simplifiedkinka}
\EN 
The solution of these differential equations provides the sought asymptotic approaches to the two vacua 
\begin{eqnarray}
\varphi(x) & = & \varphi^{(0)}_a + e^{\omega_a x} + \sum_{n=2}^\infty \mu^{(a)}_n \,e^{n \omega_a x} \,\,\,\,\,\,\,\,,\,\,\,\,\, x \rightarrow - \infty 
\label{asymkinka}\\
\varphi(x) & = & \varphi^{(0)}_b - e^{-\omega_b x} + \sum_{n=2}^\infty \mu^{(b)}_n \,e^{n \omega_b x} \,\,\,\,\,\,,\,\,\,\,\, x \rightarrow + \infty 
 \label{asymkinkb}
\end{eqnarray}
where the coefficients $\mu^{(i)}_n$'s can be iteratively computed in terms of the $\alpha^{(i)}_n$'s in (\ref{simplifiedkinka}). Notice that all the exponents of the exponentials are expressed in terms of integer multiples of the curvature $\omega_i$ of the corresponding vacuum. When $\omega_a \neq \omega_b$, the exponential approaches to the two vacua are of course different and in this case the kink configuration $\varphi_{ab}(x)$ may be not related in a simple way to the anti-kink configuration $\varphi_{ba}(x)$, say by a parity transformation. As discussed later in this paper, this 
may have far-reaching consequences on the spectrum of the theory.   
\begin{figure}[t]
\psfig{figure=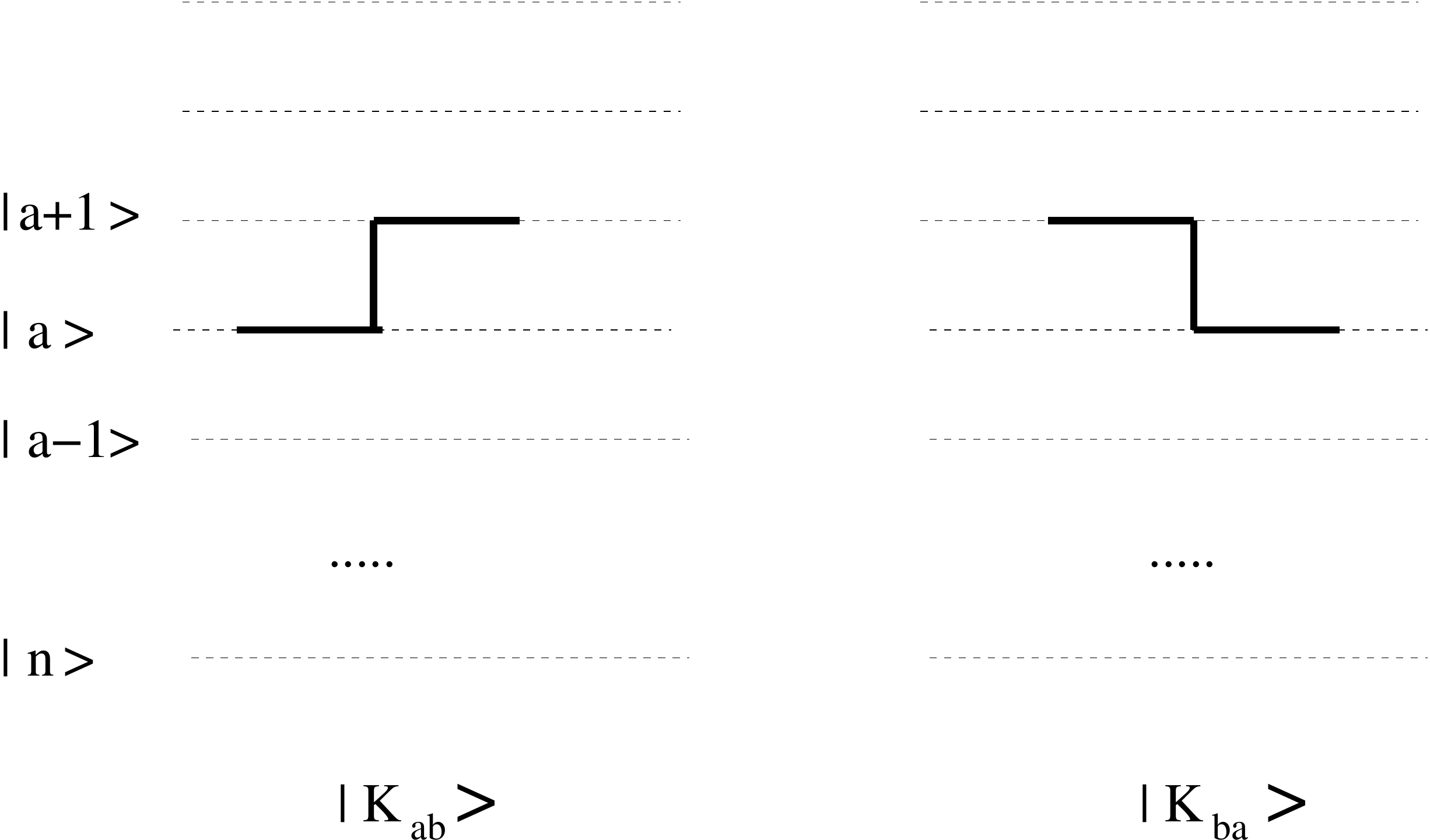,height=5cm,width=8cm}
\caption{{\em Kink and antikink configurations.}}
\label{step}
\end{figure}
\noindent

\item Associated to each solution $\varphi_{ab}(x)$ there is a classical energy density given by  
\EQ
\epsilon_{ab}(x) \,=\,\frac{1}{2} \left(\frac{d\varphi_{ab}}{d x}\right)^2 + U(\varphi_{ab}(x))  \,\,\,, 
\EN 
which, in light of (\ref{kinkequation}), can be written as 
\EQ
\epsilon_{ab}(x) \,=\,2 U(\varphi_{ab}(x)) \,\,\,. 
\EN
Integrating this density on the entire real axis, we get the classical value of the masses of this excitation    
\EQ
M_{ab} \,=\,\int_{-\infty}^{\infty} \epsilon_{ab}(x)\,dx \,\,\,.
\label{integralmass}
\EN
Notice that one does not need to know the exact solution of the kinks to get these values. Indeed, the masses $M_{ab}$ can be computed in terms of the potential $U(\varphi)$ alone: using the monotonic behavior of the kink solution, making a change of variable $t = \varphi_{ab}(x)$ and using once again eq.\,(\ref{kinkequation}), we have in fact 
\EQ
M_{ab}\,=\,\int_{\varphi_a}^{\varphi_b} \sqrt{2 \,U(\varphi)} \, d\varphi\,\,\,.
\label{finalformulamass}
\EN 
Generally the masses $M_{ab}$ of the kinks is a non-perturbative quantity of the coupling constant $\lambda$ of the theory, which diverges when $\lambda \rightarrow 0$. Hereafter a semi-classical regime has meant to be as the regime where the coupling constant is sufficiently small, so that the masses of the kinks are higher than any other mass scale. Eq.\,(\ref{finalformulamass}) shows that, even though the kink $\varphi_{ab}(x)$ and the anti-kink $\varphi_{ba}(x)$ may not be related one to the other in a simple way, their mass is nevertheless always the same. 
\item 
Using the relativistic invariance of lagrangian theory (\ref{Lagrangian}), we can set the static solutions $\varphi_{ab}(x)$ in motion by a Lorentz transformation  $\varphi_{ab}(x) \rightarrow \varphi_{ab}\left[(x \pm v t)/\sqrt{1-v^2}\right]$: these new configurations correspond to lumps of energy 
fulfilling a relativistic dispersion relation and localised at the maximum of the energy density $\epsilon_{ab}(x)$. We can then associate to them 
the quantum {\em kink states} $\mid K_{ab}(\theta)\,\rangle$ \citep{DHN,GJ,raj}, where $\theta$ is the rapidity variable which parameterises their relativistic dispersion relation  
\EQ
E = M_{ab}\,\cosh\theta
\,\,\,\,\,\,\,
,
\,\,\,\,\,\,\,
P = M_{ab} \,\sinh\theta
\,\,\,.
\label{rapidity}
\EN 
At the quantum level, the masses $M_{ab}$ of the various kinks will experience a {\em finite renormalization} with respect to their classical values (\ref{finalformulamass}) (see \citep{ DHN,raj}) and we will take explicitly into account this fact in some of the examples discussed in the sequel of the paper. 

\item The various kink configurations can be conveniently represented in a graphical way as in Figure \ref{multikinkconf}. Even though this graphical representation is a very crude image of the actual exponential approaches to the two vacua (which, as we saw, could be also different), it provides 
nevertheless a useful rule of thumb to build up and depict the multi-kink states of the quantum theory:  these are given by a string of kinks, satisfying 
the adjacency condition of the consecutive indices for the continuity of the field configuration, as shown in Figure \ref{multikinkconf}. 
\EQ
\mid K_{a_1,a_2}(\theta_1) \,K_{a_2,a_3}(\theta_2)\,K_{a_3,a_4}(\theta_3) \ldots \rangle 
\,\,\,\,\,\,\,\,
,
\,\,\,\,\,\,\,\, (a_{i+1} = a_i \pm 1)
\EN 
Notice that the field theory (\ref{Lagrangian}) has an associated conserved current $j^\mu(x) \,=\, {\cal N}\,\epsilon^{\mu\nu}\partial_\nu \varphi(x)$,  whose charge 
\EQ
Q \,=\, {\cal N} \int_{-\infty}^{\infty} \partial_1 \varphi(x) \,=\,{\cal N} \left[\varphi(+\infty) - \varphi(-\infty))\right] \,\,\,, 
\label{topologicalcharge0}
\EN 
assumes integer number values, provided we have properly chosen the constant ${\cal N}$ in reference to the various values of the vacua $\varphi_n^{(0)}$.  These integers $Q$ label the different topological sectors of the theory, where $Q = 0$ refers to the Hilbert space build upon the various vacua $\mid {\bf a} \rangle$, while $Q =\pm 1$ to the sectors of kink and anti-kink. Other values of $Q$ refer to higher multi-kink sectors \citep{raj}.

\begin{figure}[t]
\vspace{8mm}
\psfig{figure=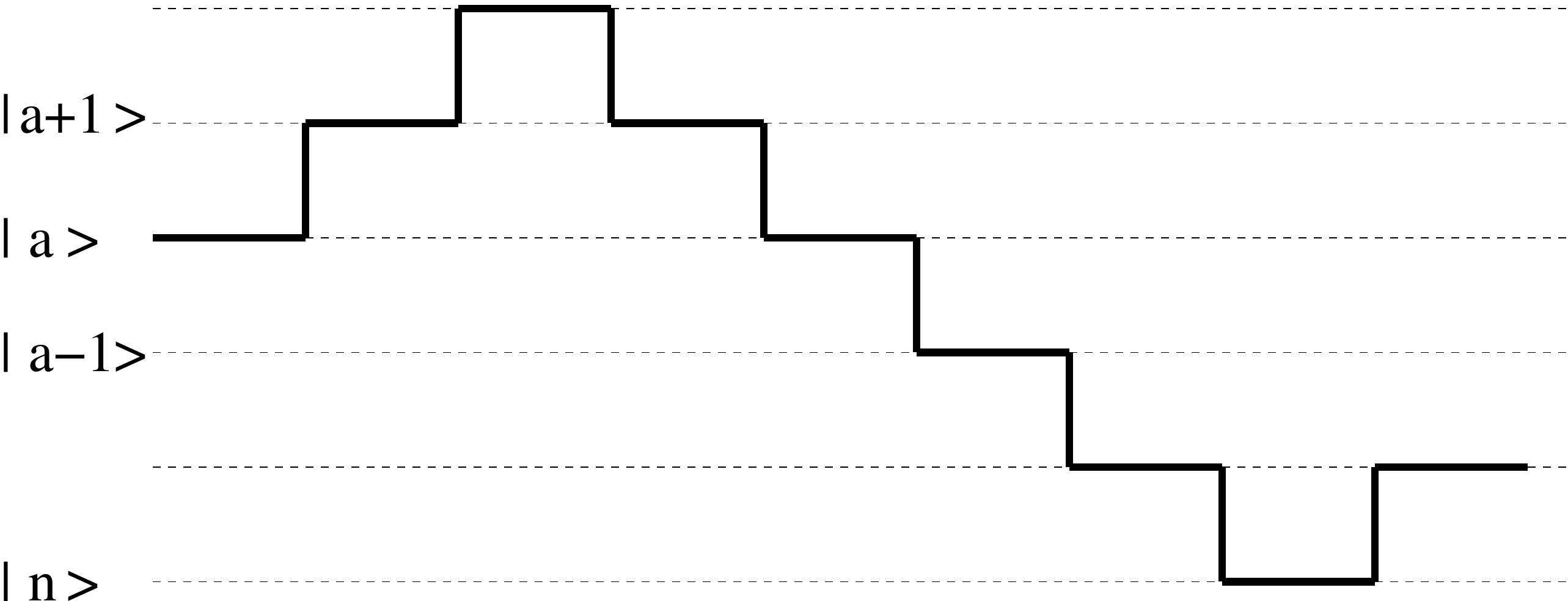,height=4cm,width=8cm}
\caption{{\em Multi-kink configurations.}}
\label{multikinkconf}
\end{figure}
\noindent
\item Important dynamical data are the bound states made of the kink-antikink excitations, alias the scalar excitations $\mid B_c(\theta) \rangle_a$, called {\em breathers}, which may exist on top of each vacua  $\mid {\bf a}\,\rangle$. These excitations live in the $Q=0$ sector of the theory.  
For a QFT based on a single real field, these breather are all non-degenerate states and, in presence of a charge conjugation 
symmetry ${\cal C}$, they may be also classified in terms of the even or odd eigenvalues of ${\cal C}$. On a general 
ground it may be argued that all scalar excitations that live on top of the vacuum $\mid {\bf a} \rangle$ must be identified as the 
bound states of the kink-antikink configurations that start and end at the same vacuum 
$\mid {\bf a}\,\rangle$, i.e. $ \mid K_{ab}(\theta_1) \,K_{ba}(\theta_2)\,\rangle$. These bound states correspond to the ``tooth'' configurations shown in Figure\,\,\ref{tooth}. Notice that if the vacuum $\mid {\bf a} \rangle$ is connected to {\em two} different vacua rather than only one,  
there are {\em two} different kink-antikink states starting and ending at $\mid {\bf a}\,\rangle$, alias $\mid K_{a,a\pm 1} K_{a\pm 1, a}\rangle$. 
However, as clarified in \citep{GMneutral}, the only one that matters for the spectrum of the breathers is the configuration with the 
{\em lower} kink mass between the two values $M_{a,a\pm1}$ or, in presence of a degeneracy of $\mid K_{a,a\pm 1} K_{a\pm 1, a}\rangle$, a special linar combination of them. This is the content of the rule of \textquotedblleft {\em The importance of being small}\textquotedblright \,that we will also advocate later in this paper. So, let's assume we have selected the kink-antikink configuration $\mid K_{ab}(\theta_1) K_{ba}(\theta_2) \rangle$ associated to the lowest mass $M_{ab}^*$ of the kink going out or coming to the vacuum $\mid {\bf a} \rangle$: we can then identify the breathers $\mid B_c\,\rangle_a$ in terms of the poles at an imaginary value $i \,u_{a b}^c$ within the physical strip $0 < {\rm Im}\, \theta < \pi$ of the rapidity difference $\theta = \theta_1 - \theta_2$ of this state
\EQ
\mid K_{ab}(\theta_1) \,K_{ba}(\theta_2) \,\rangle \,\simeq \,i\,\frac{g_{ab}^c}{\theta - i u_{ab}^c}
\,\mid B_c\,\rangle_a \,\,\,, 
\label{factorization}
\EN 
where $g_{ab}^c$ is the on-shell 3-particle coupling between the kinks and the neutral particle. Knowing the resonance value $i \,u_{ab}^c$, the mass of the bound states are simply obtained by substituing this quantity into the expression of the Mandelstam variable $s$ of the two-kink channel 
\EQ
s = 4 M^{*2}_{ab} \,\cosh^2\frac{\theta}{2} 
\,\,\,\,\,\,
\longrightarrow 
\,\,\,\,\,\, 
m_c \,=\,\pm 2 M_{ab}^* \cos\frac{u_{ab}^c}{2} \,\,\,.
\label{massboundstate}
\EN 
Obviously for a bosonic spectrum we have to choose above the $+$ sign: this will not be necessarily the case for the fermionic case, as discussed later. 
The determination of these resonance values $u_{ab}^c$ and the semi-classical computation of the spectrum of these bosonic 
neutral excitations are discussed in the next section. 
\end{enumerate}

\begin{figure}[b]
\vspace{8mm}
\psfig{figure=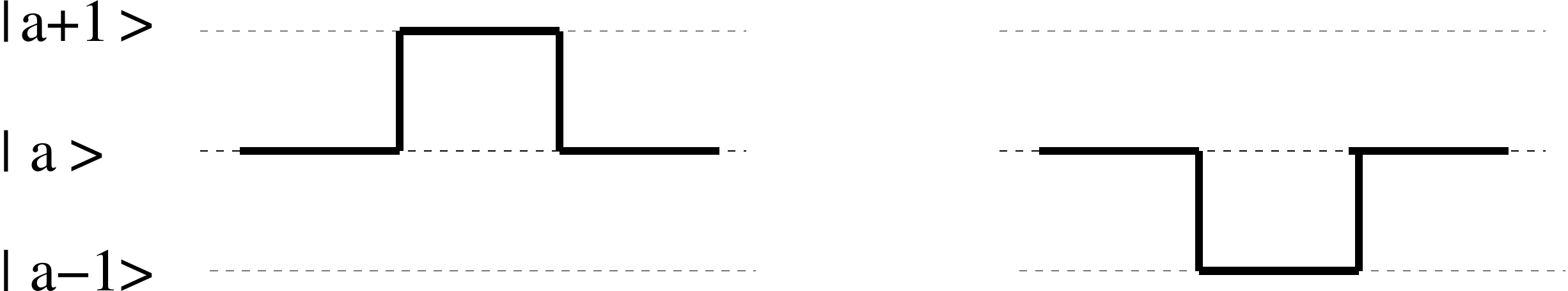,height=19.5mm,width=7.5cm}
\caption{{\em Kink-antikink configurations which may give rise to a bound state nearby the vacuum 
$\mid {\bf a}\,\rangle$.}}
\label{tooth}
\end{figure}

\section{The bosonic bound states in $Q=0$ sector}
\label{sec:semiclassboson}
\noindent
When the coupling constant $\lambda$ goes to zero, the mass of the various kinks becomes very large: in this case there is a semi-classical 
self-consistent way to compute the matrix elements on kink states of the fundamental field $\varphi(x)$. The final formula just employs the Fourier transform of the classical field $\varphi_{ab}(x)$ interpolating between the two vacua $\mid {\bf a} \rangle$ and $\mid {\bf b} \rangle$ and reads \citep{GJ,FFvolume,GMneutral}
\EQ
f_{ab}^{\varphi}(\theta) \,=\,\langle K_{ab}(\theta_2) \,\mid \varphi(0) \,\mid \, K_{ab}(\theta_1) \rangle 
\,\simeq \,\int_{-\infty}^{\infty} dx \,e^{i M_{ab} \,\theta\,x} \,\,
\varphi_{ab}(x) \,\,\,,
\label{remarkable1}
\EN  
where $\theta = \theta_1 - \theta_2$. 
This formula can be also generalised to any operator $G[\varphi(x)]$, function of $\varphi(x)$: indeed, one has simply to 
substitute above $\varphi_{ab}(x) \rightarrow G[\varphi_{ab}(x)]$. 
\begin{figure}[b]
\vspace{10mm}
\psfig{figure=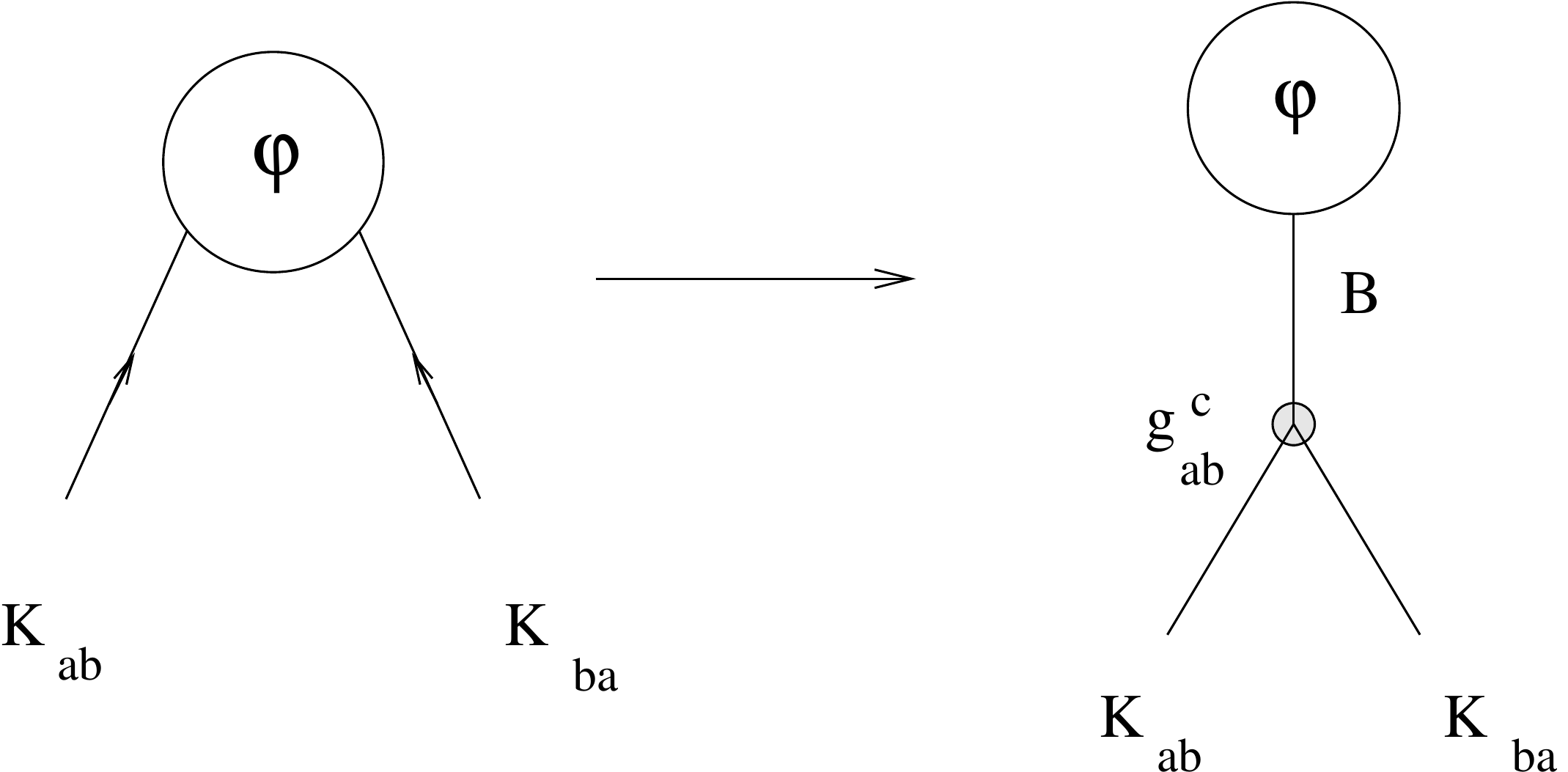,height=4cm,width=9cm}
\vspace{1mm}
\caption{{\em Graphical representation of the pole and related bound state $\mid B_c \rangle_a$ of kinks $\mid K_{ab}\rangle$ and anti-kink 
$\mid K_{ba}\rangle$.}}
\label{polematrixs}
\end{figure}
The justification and the derivation of this formula are given in the original papers \citep{GJ,FFvolume,GMneutral} (see, in particular, 
\citep{GMneutral} for its correct dependence on the relativistic invariants of the two-body channel) and will not further discussed here.

\vspace{1mm}
One can now take advantage of such a formula to determine the semi-classical spectrum of the bound states in the $Q=0$.  As we argued in 
\citep{GMneutral}, the steps to be taken are the following.  Firstly, go to the crossed channel of the matrix element by substituting in eq.(\ref{remarkable1}) $\theta \rightarrow i \pi - \theta$, so that the corresponding expression may be interpreted as the Form Factor 
\EQ
F_{ab}^{\varphi}(\theta) \,=\, f(i \pi - \theta) \,=\,\langle a \,\mid \varphi(0) \,\mid \,K_{ab}(\theta_1) \,
K_{ba}(\theta_2) \rangle \,\,\,, 
\label{remarkable2}
\EN    
where it appears the $Q=0$ neutral kink states around the vacuum $\mid {\bf a} \rangle$ that we are interested in. 
Secondly, identify the poles of $F_{ab}^\varphi(\theta)$ localised in the physical strip $0 < {\rm Im}\, \theta < \pi$, see Figure \ref{polematrixs}. 
Since the poles of a Fourier transform of a function are given in terms of the exponential asymptotic behavior of the function itself, to get the spectrum above the vacuum $\mid {\bf a} \rangle$ we have then simply to know the exponential approach of $\varphi_{ab}(x)$ to $\varphi^{(0)}_a$ for $x \rightarrow - \infty$. This asymptotic 
behavior is in eq.\,(\ref{asymkinka}): given that it consists of a series made of multiples of the same exponential (expressed by the curvature $\omega_a$ at that minimum), the poles of  $F_{ab}^{\varphi}(\theta)$ are regularly spaced by 
\EQ
\xi_a \equiv 
\frac{\omega_a}{\pi M_{ab}} \,\,\,.
\label{xia}
\EN
Focusing the attention on the kink of lower mass starting from the vacuum $\mid {\bf a} \rangle$, the semi-classical mass spectrum of the bosonic neutral bound states around the vacuum $\mid {\bf a} \rangle$ assumes then the universal form 
\EQ
m_{a}^B(n) \,=\,2 M_{ab}^* \,\sin\left(n\,\frac{\pi\,\xi_a}{2}\right) 
\hspace{5mm} , \hspace{5mm} n =1,2,\ldots N^B_a
\label{universalmassformula}
\EN 
where $N^B_a$ is the total number of bound state on top of it: this is the number of poles that fall within the physical strip and it is given by 
\EQ
N^B_a \,=\, \left[\frac{1}{\xi_a}\right] \,\,\,, 
\label{numberbosonicboundstate}
\EN
where $[x]$ expresses the integer part of the number $x$. Posing 
\EQ
m_1^B\,=\, m_{a}^B(1) \,=\, 2 M_{ab} \,\sin\left(\frac{\pi\,\xi_a}{2}\right) \,\,\,,
\EN
the formula (\ref{universalmassformula}) can be equivalently written as 
\EQ
m_{a}^B(n) \,=\,m_{1}^B \,\,\frac{\sin\left(n\,\frac{\pi\,\xi_a}{2}\right)}{\sin\left(\,\frac{\pi\,\xi_a}{2}\right)} 
\hspace{5mm} , \hspace{5mm} n =1,2,\ldots N^B_a
\,\,\,.
\label{universalmassformula2}
\EN 
Notice that, if $\xi_a > 1$ the pole(s) of $F_{ab}^\varphi(\theta)$ are outside the physical strip and there is no single bound state; in the two intervals  $\frac{1}{2} < \xi_a < 1$ and $\frac{1}{3} < \xi_a < \frac{1}{2}$ , there are one and two bound states, respectively; while for $\xi_a < \frac{1}{3}$, when the theory is non-integrable, out of the possible $[1\xi_a]$ bound states only the lowest two will be stable, the others being resonances 
\citep{GMneutral}. Indeed, for a geometrical property of the sine of multiple angles, we have $m_n > 2 m_1$ (for $n > 3$) and therefore 
these higher particles are generically expected to decay. Hence, the conclusion is that, in a generic non-integrable bosonic theory, each vacuum cannot have more than two stable neutral particles above it.  

\section{Fermion in a bosonic background}
\label{sec:semiclassfermion}
\noindent
Let's now consider a two-dimensional theory involving a scalar field $\varphi(x)$ and a fermion field 
\EQ
\psi(x) \,=\,
\left(\begin{array}{c} \psi_1 \\ \psi_2 \end{array} \right)
\label{fermionbid}
\EN 
with Lagrangian density 
\EQ
{\mathcal L}\,=\, \frac{1}{2} (\partial_{\mu} \varphi)^2 - U(\varphi,\lambda) \,\,\, + i \bar\psi \gamma^{\mu}\,\partial_{\mu} \,\psi  - 
V(\varphi, g) \, \bar\psi \,\psi \,\,\,. 
\label{Lagrangianfermion}
\EN 
As coupling constants we have now both $\lambda$ and $g$. In addition to the bosonic potential $U(\varphi,\lambda)$ (that we will take 
as those in Section 2,  alias with a degenerate set of minima connected by the kinks $\varphi_{ab}(x)$),  we have now a Yukawa interaction $V(\varphi)$ 
for the fermion. For this interaction we require that it satisfies the following conditions for any kink configuration $\varphi_{ab}(x)$ 
\EQ
V(\varphi_{ab}(x), g) \,=\,
\left\{
\begin{array}{l}
v_a(g)  \,\,\,\,\,, \,\,\,\,\, x \rightarrow - \infty \\
v_b(g)  \,\,\,\,\,, \,\,\,\,\, x \rightarrow + \infty
\end{array}
\right.
\label{twolimits}
\EN
where both limits are finite and such that their product is always strictly negative 
\EQ
v_a(g) \, v_b(g) \, < 1 \,\,\, .
\label{negativitycondition}
\EN 
From now on, in order to simplify the notation, we will denote by $V_{ab}(x)$ the potential $V(\varphi, g)$ computed on the 
kink configuration $\varphi_{ab}(x)$ and, for simplicity, we will also skip the presence of the coupling constant $g$ in this expression. 

For the $\gamma$ matrices we adopt the Weyl representation given by 
\EQ
\gamma^0 \,=\,\sigma_2 \,=\,\left(
\begin{array}{rr}
0 & - i \\
i & 0 
\end{array} 
\right)
\hspace{3mm}
;
\hspace{3mm}
\gamma^1 = i \sigma_1 \,=\,\left(
\begin{array}{rr}
0 &  i \\
i & 0 
\end{array} 
\right)
\hspace{3mm}
,
\hspace{3mm}
\gamma^5 \,=\, \gamma^0 \gamma^1 \,=\, \sigma_3 \,=\,
\left(
\begin{array}{rr}
1 &  0 \\
0 & -1
\end{array} 
\right)
\,\,\,.
\EN
The charge-conjugation matrix $C$, satisfying 
\[
(C \gamma^0) \,(\gamma^{\mu})^* \, (C \gamma^0)^{-1} \, =\, - \gamma^\mu \,\,\,,
\] 
is given by $C = \gamma^0$ and maps the fermion $\psi$ to its conjugate particle $\psi_c$ according to 
\EQ
\psi_c \,=\, (C \gamma^0) \psi^* \,\,\,.
\label{conjugationfermion}
\EN
In this representation a Majorana fermion (which satisfies the neutrality condition $\psi_c \,=\,\psi$) has then both components real 
\EQ
\psi_1^* \,=\, \psi_1 
\hspace{5mm}
,
\hspace{5mm}
\psi_2^* \,=\, \psi_2 
\,\,\,.
\EN   
While in \citep{DHN,JR} it was considered the situation where the Lagrangian (\ref{Lagrangianfermion}) employs a 
complex Dirac fermion -- and this leads to the discovery of the fractionalization of the fermion quantum number \citep{JR} -- 
here for simplicity we are interested in studying directly the case where $\psi(x)$ is a real Majorana fermion. The two situations 
are closely related, of course, and indeed our final result for the fermionic bound state spectrum in the $Q=0$ sector, 
eq.\,(\ref{generalexpressionmassfermion}), applies equally well to both cases. However, the key difference between Dirac/Majorana 
fermions consists of the impossibility to define, in the Majorana case, a {\em fermion number} as the conserved quantity associated to the 
transformation $\psi \rightarrow e^{i \alpha} \, \psi$, simply because the Majorana fermion is real. So, in the case of Majorana fermion,  
we will simply talk about a fermion parity ${\cal P}$. A field theory similar to the one discussed in this paper was also considered in \citep{Stern}.   

Notice that the Lagrangian (\ref{Lagrangianfermion}) has no explicit fermion mass term and therefore the fermion looks massless. Yet 
the problem addressed in this paper is precisely how to find the fermionic mass spectrum in the $Q=0$ topological sector. As a very preliminary 
observation in this direction, it is clear that it will be the interaction term $V(\varphi(x))$ to play the role of an effective mass for the fermion field $\psi(x)$ and this is particularly true when the bosonic field $\varphi(x)$ has been prepared in a $Q=0$ trivial state, such as one of its vacua: in these cases, the the effective mass of the fermion at the vacuum $\mid {\bf a} \rangle$ is given by the constant value $V(\varphi_a^{(0)}) = v_a$. However, as we saw it was the case for the bosonic spectrum, to unveil the actual spectrum of the fermion in the $Q=0$ sector (and see, in particular, whether the particle associated to $\psi(x)$ belongs or not to the physical spectrum!) we need to study first the higher topological sectors of the theory. 

\begin{figure}[t]
\vspace{10mm}
\psfig{figure=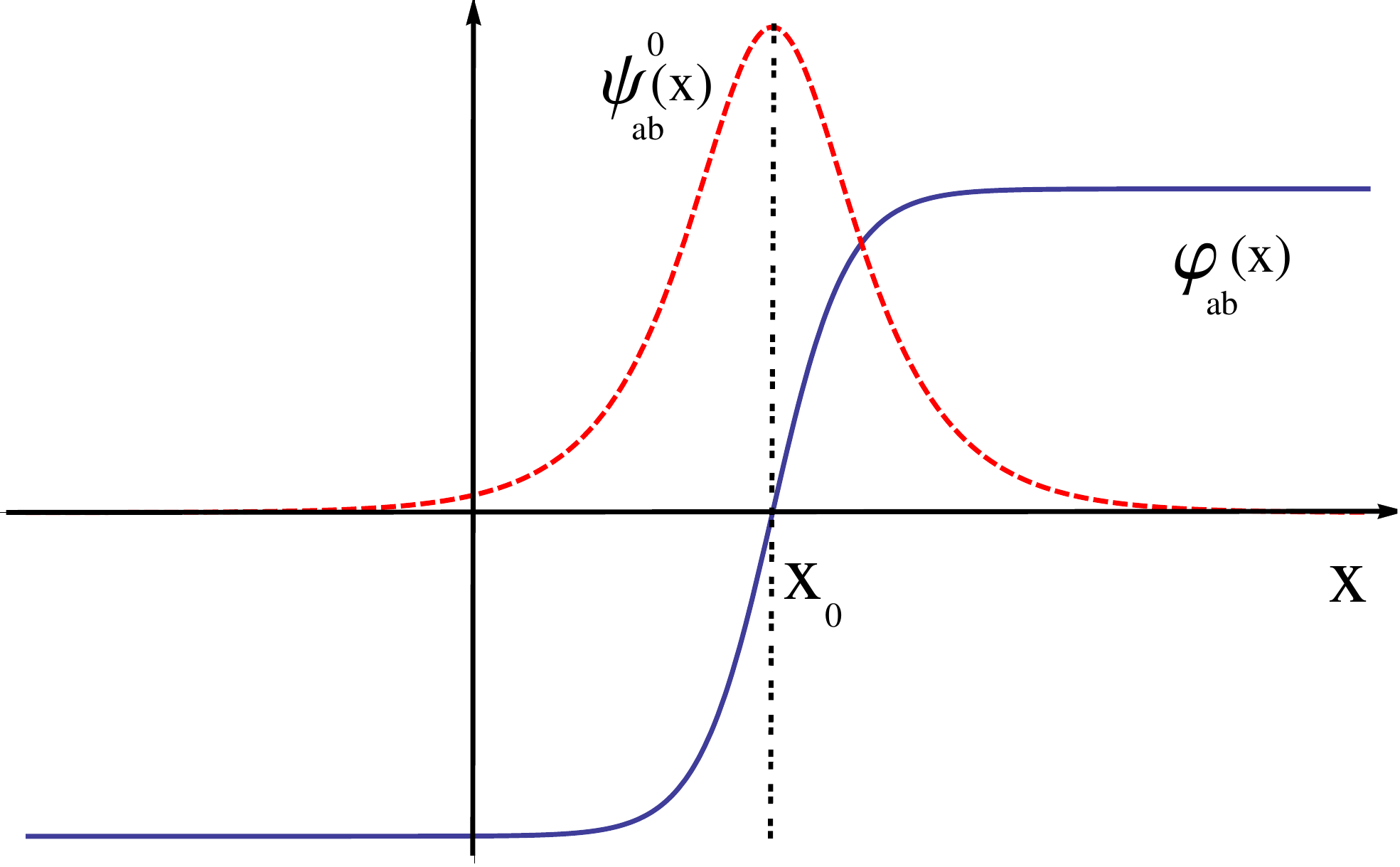,height=4cm,width=7cm}
\vspace{1mm}
\caption{{\em Kink configuration $\varphi_{ab}(x)$ and the associated fermion zero mode $\psi_{ab}^{(0)}(x)$. The latter is always localised at the point $x_0$ where the former makes a transition between the two vacua.}} 
\label{zeromodefigure}
\end{figure}

Let's consider then the $Q = 1$ sector (analogous considerations hold for $Q = -1$), with the bosonic field prepared 
in the one kink static configurations $\varphi_{ab}(x)$. Although, strictly speaking,  such a classical configuration refers to the equation 
of motion (\ref{static}) where the fermion field is absent, yet  it is natural to assume (as we will do) that the latter will have 
a negligible effect on the former as far as the mass of the kink itself far exceeds the \textquotedblleft mass energy\textquotedblright  
of the fermion for both vacua, i.e. if 
\EQ
M_{ab} \gg \mid v_i \mid 
\,\,\,\,\,
,
\,\,\,\,\,
i = a, b \,\,\,. 
\label{conditionssemi}
\EN
This will be the condition expressing the semi-classical limit of the Lagrangian (\ref{Lagrangianfermion}). In the $Q=1$ sector the fermion has a new qualitative non-perturbative feature: it possesses localised zero-energy mode, namely a normalizable solution of zero energy of  the Dirac equation \citep{DHN,JR}. Indeed, being $\varphi_{ab}(x)$ a {\em static} configuration, we can as well as look for a {\em static} solution of the Dirac equations, which therefore take the form
\EQ
\begin{array}{l}
\partial_x \psi_2^{(0)} + V_{ab}(x) \, \psi_1^{(0)} \,=\,0 \,\,\,,\\
\partial_x \psi_1^{(0)} + V_{ab}(x) \, \psi_2^{(0)} \,=\,0 \,\,\,.
\end{array}
\label{Diracequation}
\EN 
These are nothing else but the $0$-eigenvalue condition $H_D \, \psi^{(0)} \,=\, 0$
for the Dirac Hamiltonian 
\EQ
H_D \,=\, - i \alpha \partial_x + V_{ab}(x) \, \beta \,\,\,,
\label{DIRACHAM}
\EN
where $\alpha = \gamma^0 \gamma^1 = \sigma_3$ and $\beta = \gamma^0$. Solutions of the two equations (\ref{Diracequation}) 
can be found in terms of the linear combinations of the two components of the fermion 
\EQ
\psi_{\pm}^{(0)} (x)\equiv \left(\psi_1^{(0)} \pm \psi_2^{(0)} \right) \,=\, 
A_{\pm} \, \exp\left[
\mp \,\int_{x_0}^x V_{ab}(t) \, dt \right] 
\,\,\,.
\label{solutionnn}
\EN  
In light of the asymptotic behaviors (\ref{negativitycondition}) of the potential $V_{ab}(t)$ and the condition (\ref{negativitycondition}) 
on their signs, it is clear that only one of the two solutions $\psi_{\pm}^{(0)}(x)$ is normalizable, the other being divergent at infinite. To fix the ideas, 
let's assume that $v_a < 0$ and $v_b >0$: in this case, the normalizable solution is $\psi_+(x)$ and correspondingly we have to impose $A_- = 0$,  
so that the final result for wave function of the zero mode is  
\EQ
\psi_{ab}^{(0)}(x) \,=\,A \,\hat\psi\, \, 
\exp\left[- \,\int_{x_0}^x V_{ab}(t) \, dt \right] \,\,\,,
\label{zeromode}
\EN 
where $\hat\psi = \left(\begin{array}{c} \,\,\,1 \\ -1 \end{array} \right)$ and $A$ is a normalization constant. This wave function is usually 
localised at the position where the kink solution jumps between the two vacua, as shown in Figure \ref{zeromodefigure}.

\vspace{1mm}
In the $Q=1$ sector, the quantization of the fermion field is achieved by its eigenmode expansion \citep{DHN,JR}
\EQ
\psi(x,t) \,=\, 
a_0 \, \psi_{ab}^{(0)}(x) + \sum_{p} \left[b_p \,e^{-i E_p t} \, \psi_{E_p}(x) + b_p^\dagger \, e^{i E_p t}\, \psi_{-E_p}(x) \right] \,\,\,,
\label{fermionexpansion}
\EN 
where the operator $a_0$ is relative to the zero-mode solution, while $b_p$ and $b^\dagger_p$ are the operators associated to the positive/negative energy solutions $\psi_{\pm E}(x)$ of the Dirac Hamiltonian in the kink background $\varphi_{ab}(x)$ 
\EQ
H_D \, \psi_{\pm E}(x) \,=\, \pm E \, \psi_{\pm E}(x)\,\,\,.
\label{DiracposnegE}
\EN
The operators $b_p$ and $b_p^\dagger$ satisfy the anti-commutation relations
\EQ
\{b_p, b^\dagger_k\} \,=\,\delta( p - k) \hspace{5mm} , \hspace{5mm} \{b_p,a_0\},\,=\,\{b^\dagger_p,a_0\} = 0\,\,\,. 
\EN 
The operator $a_0$ is instead a Majorana self-conjugated zero mode operator, $a_0 = a_0^\dagger$ that, properly normalised, fulfills the condition 
\EQ
a_0^2 \,=\, 1  \,\,\,.
\label{squarezeromode}
\EN
\begin{figure}[b]
\psfig{figure=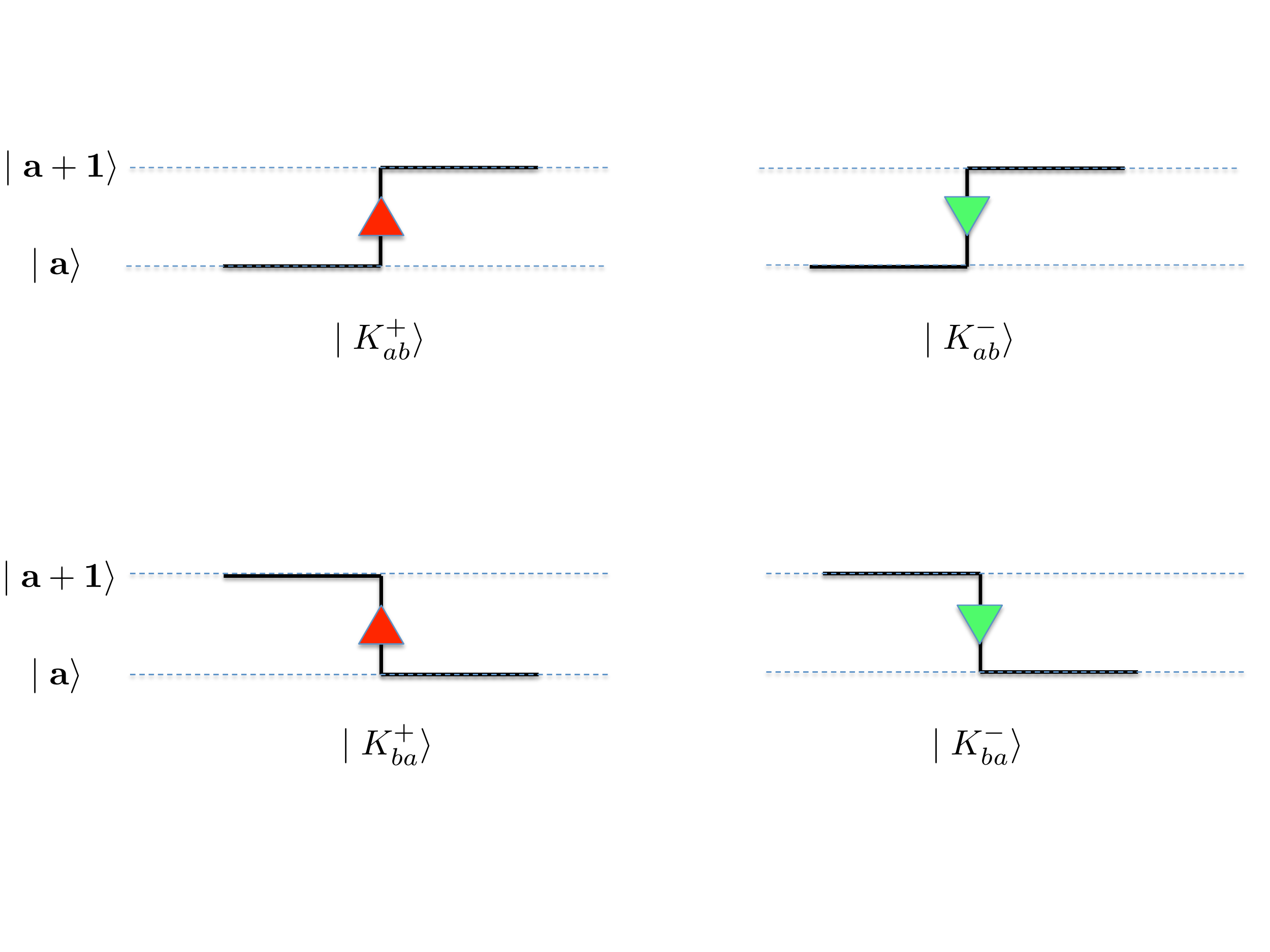,height=8cm,width=12cm}
\caption{{\em Graphical representation of the fermionic kinks $\mid K^\pm_{ab} \rangle$ and anti-kink $\mid K^\pm_{ba} \rangle$.
The red up arrow denotes those of fermion parity  ${\cal P}= +1$, while the down green arrow those of fermion parity ${\cal P}= -1$.}}
\label{kinkMajorana}
\end{figure}
The $a_0$ mode carries zero energy and therefore the original quantum state $\mid K_{ab}(\theta)\rangle$ has to be actually doubly degenerate \citep{JR}: we will use the notation $\mid K_{ab}^\pm(\theta)\rangle$ to identify these two degenerate states, related one to the other by 
\EQ
\begin{array}{l}
 \mid K_{ab}^+(\theta) \rangle \,=\, a_0 \, \mid K_{ab}^-(\theta) \rangle \,\,\,,\\
 \\
 \mid K_{ab}^-(\theta) \rangle \,=\, a_0 \, \mid K_{ab}^+(\theta) \rangle \,\,\,.
\end{array}
\,\,\,
\label{thetwokinks}
\EN
Notice that we can consider $\mid K_{ab}^+ \rangle$ to be the charge-conjugated state of $\mid K_{ab}^- \rangle$ and viceversa. 
Given their symmetry in sharing the Majorana fermion, there is a certain arbitrariness in assigning the fermion parity: our choice is 
regard $\mid K_{ab}^\pm(\theta) \rangle$ as kinks that have Majorana fermion parity ${\cal P} = \pm 1$. 
The same considerations apply to $Q = -1$ sector, alias the sector of the anti-kinks $\mid K_{ba}\rangle$, which now become doublets $\mid K_{ba}^\pm(\theta) \rangle$. We will graphically distinguish these new kinks by putting on their previous graphical representation a red up arrow on the those of fermion parity $+1$ while a green down arrow on those of fermion number $- 1$, as shown in Figure \ref{kinkMajorana}. 

\begin{figure}[t]
\psfig{figure=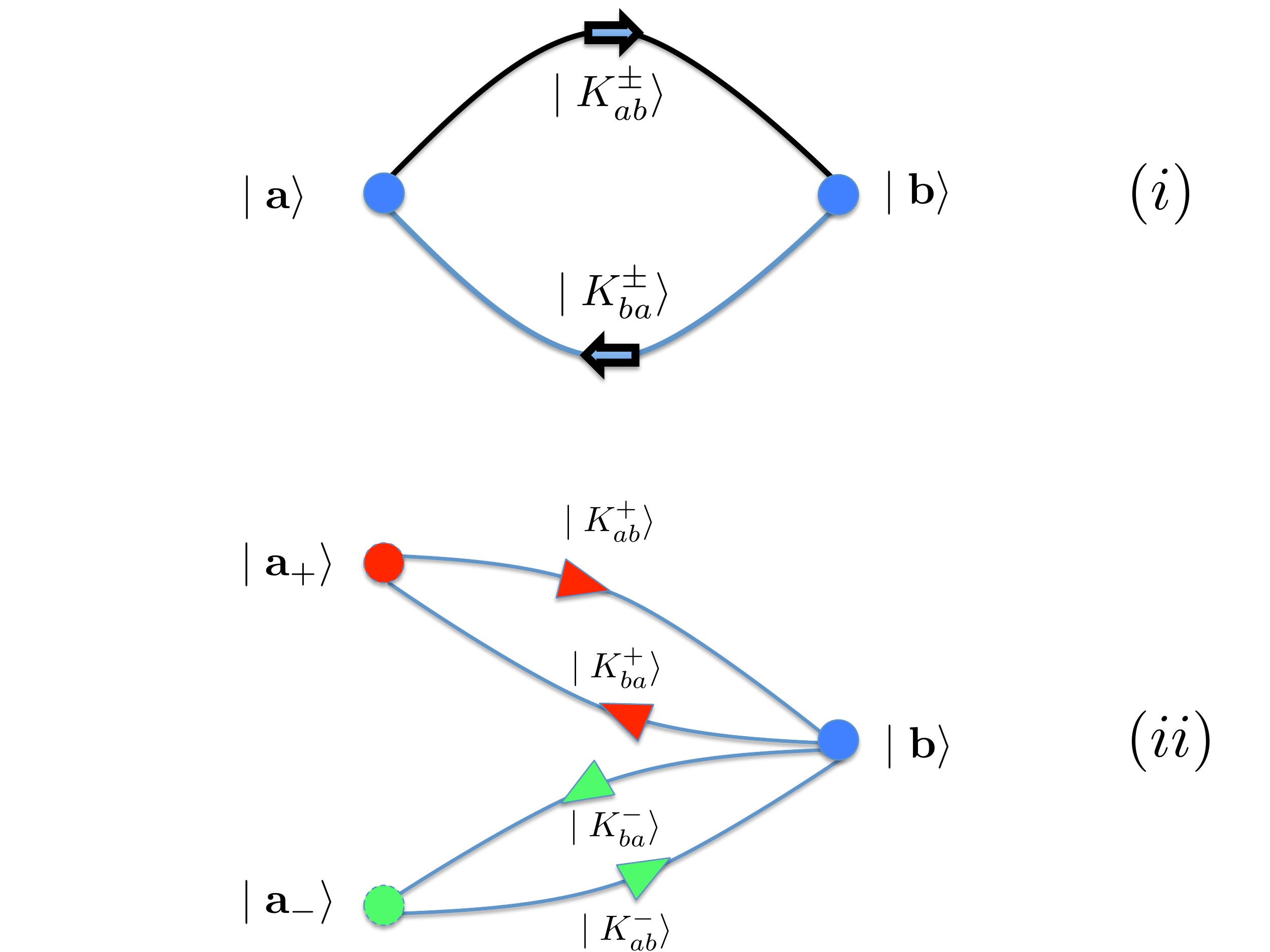,height=8cm,width=9cm}
\vspace{1mm}
\caption{{ \em (i) Adjacency diagrams of original vacua connected by fermionic kinks; (ii) splitting the original vacuum $\mid {\bf a}\rangle$ into two vacua $\mid {\bf a_{\pm}}\rangle$, where each of them is related to the vacuum $\mid {\bf b}\rangle$ by kinks of the same Majorana fermion parity.}} 
\label{adjacency}
\end{figure}

\vspace{1mm}
We can now use this double multiplicity of kinks and antikinks to {\em unfold} the original vacuum structure. Namely, consider two bosonic 
neighborhood vacua $\mid{\bf a} \rangle$ and $\mid {\bf b} \rangle$ connected by the kinks and antikinks of  
different fermion number; then, the adjacency properties of the initial vacua, carried by the kinks $\mid K_{ab}^\pm\rangle$, can 
be equivalently expressed as the adjacency conditions of a new set of vacua, in which we {\em split} one of the original vacua (say $\mid {\bf a} \rangle$) into two new ones $\mid {\bf a_\pm} \rangle$, while we leave the other $\mid {\bf b} \rangle$ untouched, and we connect this new set of vacua as follows: $\mid {\bf a_+} \rangle$ is linked to $\mid {\bf b}\rangle$ only by kink/antikink of fermion parity $+ 1$, while  $\mid {\bf a_-} \rangle$ is connected to $\mid {\bf b}\rangle$ only by those of fermion parity $ - 1$, see Figure \ref{adjacency}. This construction can be repeated for any two neighborhood vacua and will be useful later.

\vspace{1mm}
Let's close this section by briefly discussing the non-zero energy solutions of the Dirac Hamiltonian equation (\ref{DiracposnegE}) in the $Q=1$ sector: 
notice that if 
$$
\psi_{E} = \left(\begin{array}{c} \psi_\eta \\ \psi_\xi\end{array}\right)
$$ 
stays for a positive energy solution, then the solution corresponding to the negative energy is simply given by 
$$
\psi_{-E} = \left(\begin{array}{c} \psi_\xi \\ \psi_\eta\end{array}\right)\,\,\,,
$$ 
alias there is always a pairing between the positive/energy Hilbert space. Moreover, the eigenvalue problem (\ref{DiracposnegE}) can be equivalently expressed in terms of two Schr\"{o}dinger equations 
\EQ
\begin{array}{l}
\left[ -\partial_x^2 + (V_{ab}^2 + V_{ab}') \right]\, u \,=\, E^2 \,u \,\,\, ; \\
\left[ -\partial_x^2 + (V_{ab}^2 - V_{ab}') \right]\, v \,=\, E^2 \,v \,\,\, ,  
\end{array}
\label{SchroSUSY}
\EN 
for the linear combinations of the two components of these spinors  
\[
\begin{array}{l}
u \,=\, \psi_\eta + \psi_\xi \,\,\,,\\
v \,=\, \psi_\eta - \psi_\xi \,\,\,.
\end{array}
\]
It is worth noticing that the equations (\ref{SchroSUSY}) are those of one-dimensional Supersymmetry Quantum Mechanics 
and shape-invariant potentials \citep{Witten,Khare}, an observation that may be quit useful to find an explicit form of eigenvalues and eigenvectors in the $Q=\pm 1$ sectors (see, for instance, \citep{iranian}). 

\section{The fermionic bound states in $Q=0$ sector}\label{fermionsemiclfor}
\begin{figure}[t]
\psfig{figure=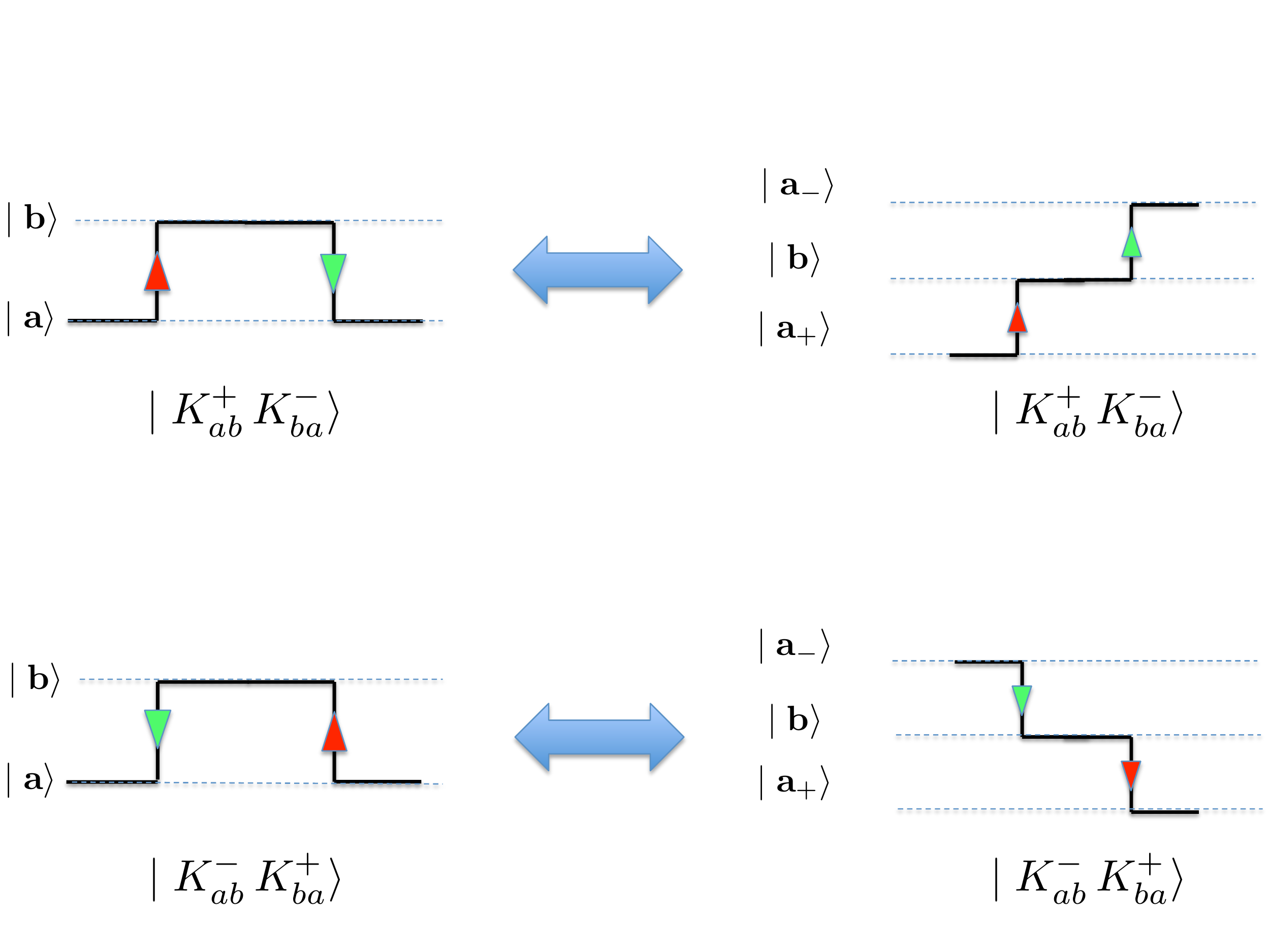,height=8cm,width=12cm}
\vspace{1mm}
\caption{{\em Two-particle fermionic kinks in the graphical representation of the original vacua (figures on the left) and of the unfolded vacua (figures on the right).}} \label{unfoldedkink}
\end{figure}

\noindent
Let's now study the fermionic bound states spectrum in the $Q=0$ sector. To this aim, it is useful to remind that in the paper \citep{JR} 
Jackiw and Rebbi derived semi-classical matrix elements of the fermion field $\psi(x)$ on the kink states $\mid K_{ab}^\pm(\theta)\rangle$. 
Actually, as they are written in \citep{JR}, these matrix elements do not take into a proper account the dependence on the relativistic invariants of the channel of the two kinks. But this is a problem that can be easily cured, similarly to what was already done for the pure bosonic case in \citep{FFvolume,GMneutral}: one has simply to introduce the rapidity variable $\theta$ and consider that, in semi-classical approximation, $M_{ab} \sinh\theta \simeq M_{ab} \theta$, given that the kinks must be regarded to be almost static being their mass very large. In this way, the definite semi-classical formula for the fermion matrix elements on the kink states reads as  
\EQ
g_{ab}^+(\theta) \,=\, g_{ab}^-(\theta) \,\simeq\,\int_{-\infty}^{\infty} dx e^{i M_{ab} \theta x} \, \psi_{ab}^{(0)}(x) \,\,\,,
\label{SEMIFERMION}
\EN 
where we have put
\EQ
g_{ab}^{\pm}(\theta) \,=\, \langle K_{ab}^\mp(\theta_1) \mid \psi(0) \mid K_{ab}^\pm(\theta_2) \rangle \,\,\,.
\EN 
In eq.\,(\ref{SEMIFERMION}), $\theta = (\theta_1 - \theta_2)$ and $\psi_{ab}^{(0)}(x)$ is the zero-mode configuration of the fermion in the presence 
of the kink background $\varphi_{ab}(x)$. The identity of the two matrix elements $g_{ab}^\pm(\theta)$ is due to the real condition on the Majorana fermion $\psi(x)$. For a Dirac fermion field these matrix elements are complex conjugate one to the other and eq.\,(\ref{SEMIFERMION}) must be 
substituted instead by the identity 
\EQ
\langle K_{ab}^\mp(\theta_1) \mid \psi(0) \mid K_{ab}^\pm(\theta_2) \rangle \,=\, 
\langle K_{ab}^\mp(\theta_2) \mid \psi^\dagger(0) \mid K_{ab}^\pm(\theta_1) \rangle^*\,\,\,.
\label{matricDiracfermion}
\EN
Having now the correct dependence of this matrix element on the relativistic invariants of the channel of the two kinks, as in the bosonic we can make the analytic continuation $\theta \rightarrow i \pi - \theta$ of (\ref{SEMIFERMION}) and go to the crossed channel
\EQ
G_{ab}^\pm(\theta) \,=\,g_{ab}^\pm(i \pi - \theta) \,=\,\langle a \mid \psi(0) \mid K_{ab}^\pm(\theta_1) K_{ba}^\mp(\theta_2) \rangle \,\,\,.
\EN 
Therefore, for the $Q=0$ bound states with fermion quantum number around the vacuum $\mid {\bf a} \rangle$, 
we have to look at the poles of $G_{ab}^\pm(\theta) $ in the physical strip $0 < {\rm Im}\, \theta < \pi$. Notice that there are {\em two} states 
which will share the same poles, alias $\mid K_{ab}^+ K_{ba}^- \rangle$ and $\mid K_{ab}^- K_{ba}^+ \rangle$: these states are 
charge-conjugated one to the other, in agreement with the general result that any solution with $+E$ is paired with another of $-E$. Notice that, unfolding the original vacua, these states $\mid K_{ab}^\pm K_{ba}^\mp \rangle$ have the graphical representation shown in Figure \ref{unfoldedkink}. 

\vspace{1mm}
Given both the conditions (\ref{twolimits}) and (\ref{negativitycondition}), it is clear that zero-mode goes exponentially to zero at $x \rightarrow \pm\infty$ 
\EQ
\psi_{ab}^{(0)}(x) \,=\, A\, \hat\psi\, 
\left\{
\begin{array}{lll}
\exp\left[v_a(g) x\right] & ,& x \rightarrow - \infty \\
\exp\left[-v_b(g) x\right] & ,& x \rightarrow + \infty
\end{array}
\right.
\label{twolimitszeromode}
\EN
As in the bosonic case, for what concerns the bound states relative to the vacuum $\mid {\bf a} \rangle$, the only poles that matter are those coming from the behavior of this function (relative to the kink of lowest mass) at $x \rightarrow -\infty$: in particular, the first pole of $G_{ab}^\pm(\theta)$ is simply fixed by the asymptotic value $v_a$
\EQ
\theta_1 \,=\, i \pi \left(1 - \frac{v_a}{\pi M^*_{ab}}\right)\,\,\,. 
\EN
and it is in the physical strip if 
\EQ
\eta_a(g) \,\equiv \,\frac{v_a(g)}{\pi M^*_{ab}} < 1 \,\,\,.  
\label{primopolo}
\EN 
When this happens, we have two $Q=0$ fermionic bound states of mass 
\EQ
m_1^F(g) \,=\,\pm 2 M^*_{ab} \, \sin
\left(\frac{\pi\,\eta_a(g)}{2}\right) \,\,\,. 
\label{universalmassformulaferm}
\EN 
 Few comments on this formula: 
\begin{enumerate}
\item The $\pm$ signs refer to the paired solutions of energy $\pm E$. From now on for simplicity we will concentrate the attention only on one of them, say the positive energy bound state. 
\item
When $\eta_a < 1$, we have that $\mid m_1^F(g)\mid < 2 M^*_{ab}$, i.e. we are in presence of genuine $Q=0$ fermionic bound states 
made of the kink-antikink of lowest mass carrying Majorana fermion number.
\item
In the very deep semi-classical limit, when $M^*_{ab} \rightarrow \infty$, expanding in series the formula (\ref{universalmassformulaferm}), we have that 
$ \mid m_1 \mid \simeq v_a $, which is of course the mass usually associated to the elementary fermion $\psi$ in the $Q=0$ sector at the vacuum $\mid {\bf a} \rangle$. 
\item If, on the contrary, moving the coupling constant $g$, we reach a value $g_c$ for which $\eta_a(g_c) \geq 1$, then $\mid m_1\mid $ 
overpasses the threshold value $2 M^*_{ab}$ and leaves the physical spectrum. Therefore, similarly the purely bosonic case \citep{GMneutral}, the picture is as follows: the particle associate to the fermion present in the Lagrangian (\ref{Lagrangianfermion}) has to be considered, on a general ground, not an elementary particle but a bound state of the two kinks with different fermion parity, and it belongs to the physical spectrum of the theory if and only if the condition (\ref{primopolo}) is satisfied. 
\end{enumerate}
In order to get the other poles eventually present in the Fourier transform of $\psi_{ab}^{(0)}(x)$, we need to analyse its exponential approach at 
$x\rightarrow -\infty$. This behavior evidently depends on the interaction term $V(\varphi(x))$ and, in absence of any further information on this 
function, we cannot reach any general conclusion. However, we can pin down the general pattern of the asymptotic behavior of the zero-mode in the very important cases when $V(\varphi)$ admits a series expansion around the vacuum values $\varphi_a^{(0)}$. Indeed, assume that nearby the vacuum values $\varphi_a^{(0)}$ the interaction term can be expanded as 
\EQ
V_{ab}(\varphi(x)) \,=\,v_a + V'_{ab} (\varphi(x) - \varphi_{a}^{(0)}) + \frac{1}{2} V''_{ab} (\varphi(x) - \varphi_{a}^{(0)})^2 + \cdots 
\EN 
Substituting now in this expression the quantity $(\varphi(x) - \varphi_{a}^{(0)})$ with the exponential approach given in (\ref{asymkinka}), we see 
that all terms but the first are expressed in terms of purely exponential terms which are multiple of a single one. Namely, the expression 
of $V_{ab}(\varphi(x))$ around $x \rightarrow - \infty$ can be generally written as 
\EQ
V_{ab}(\varphi(x)) \,=\, v_a + \sum_{n=1}^\infty d_n e^{n \omega_a x} \,\,\,\,\, , \,\,\,\,\,\, x \rightarrow - \infty \,\,\,, 
\label{generalexpressionV_ab}
\EN 
where $\omega_a$ is the curvature of the {\em bosonic potential} $U(\varphi)$ at $\varphi = \varphi_a^{(0)}$, while $d_n$ are coefficients  
determined by combining both the various derivatives $V^n_{ab}$ of the interaction term and the expansion coefficients $\mu_n^{(a)}$ in  (\ref{asymkinka}). However, we need an extra step to control the fermion zero mode itself, namely we need to evaluate the integral of 
$V_{ab}(\varphi)$ for $x \rightarrow - \infty$. This can be easily done using eq.\,(\ref{generalexpressionV_ab}): indeed, disregarding 
the various constants of integration and introducing other appropriate coefficients $\hat d_{n}$, the general expression of this integral can be cast as 
\EQ
\int_{x_0}^x V_{ab}(t) \, dt \,=\, -v_a x + \sum_{n=1} \hat d_n e^{n \omega_a x}  \,\,\,\,\, , \,\,\,\,\,\, x \rightarrow - \infty \,\,\,, 
\EN 
where all terms, but the first one, exponentially small. Therefore, the final result is as follows: when $V_{ab}(\varphi)$ is analytic around 
the vacuum configurations $\varphi_a^{(0)}$, we have the following asymptotic expansion for $x \rightarrow - \infty $ of the fermionic zero mode 
\begin{eqnarray}
\psi_{ab}^{(0)}(x) & \,=\, & A \,\hat\psi\, \, \exp\left[- \,\int_{x_0}^x V_{ab}(t) \, dt \right] \,=\, A\, \hat\psi\, 
\exp\left[ -v_a x + \sum_{n=1} \hat d_n e^{n \omega_a x} \right] \\
\nonumber \\
& \,=\,& A \,\hat\psi\, \, e^{v_a x} \, \left(1 + \sum_{n=1} t_{n}\, e^{n \omega_a x}\right)  
\,\,,
\label{zeromodeexpterm}
\end{eqnarray}
where $t_n$ are the final coefficients resulting from the series expansion of the exponential function and the coefficients $\hat d_n$. 
The presence of these exponential terms induces poles in the function $G_{ab}^\psi(\theta)$ localised at 
\EQ
\theta_n \,=\,i \pi (1 - \eta_a - n \,\xi_a) \,\,\,. 
\EN
They are on the physical strip as far as $n \leq (1 - \eta_a)/\xi_a$, whose integer part is therefore the total number of fermionic bound state
in the $Q=0$ sector
\EQ
N^F_a\,=\,\left[\frac{1-\eta_a}{\xi_a}\right] \,\,\,.
\label{numberfermionicboundstate}
\EN
The general semi-classical expression of the masses of the fermionic bound states is thus given by   
\EQ
m_n^F \,=\, 2 M^*_{ab} \sin\left(\frac{\pi (\eta_a + n \xi_a)}{2}\right) 
\hspace{5mm} , \hspace{5mm} n =1,2,\ldots, N^F_a  \,\,\,, 
\label{generalexpressionmassfermion}
\EN
which can be also written as 
\EQ
m_n^F \,=\, m_1^F \, \frac{\sin\left(\frac{\pi (\eta_a + n \xi_a)}{2}\right)}{\sin\left(\frac{\pi \eta_a }{2}\right)}
\hspace{5mm} , \hspace{5mm} n =1,2,\ldots, N^F_a 
\,\,\,.
\label{generalexpressionmassfermion2}
\EN
This is the main result of the paper. Comparing this mass formula with the one of the bosonic bound states, eq.\,(\ref{universalmassformula}), we see that the only significant difference is the shift induced by the quantity $\eta_a$.  In the next sections we are going to discuss a series of interesting QFT and study the various patterns of bound states that emerge in those cases.  

\section{Symmetric Wells}
\label{symmetriccase}
\noindent
The first theory we want to discuss is the familiar symmetric double well bosonic potential that conveniently we choose to write as 
\EQ
U(\varphi) \,=\,\frac{\lambda^2}{2} \left(\varphi^2 - \frac{m^2}{2 \lambda^2}\right)^2 \,\,\,.
\EN 
We denote with $\mid \pm 1 \,\rangle$ the vacua corresponding to the classical minima $\varphi_{\pm}^{(0)} \,\equiv \pm \hat\varphi\,=\pm \frac{m}{\sqrt{2} \lambda}$. 
The kink/anti-kink solutions interpolating between them are 
\EQ
\varphi_{-a,a}(x) \,=\,a \,\hat\varphi\,\tanh\left[\frac{m x}{\sqrt{2}}\right]
\,\,\,\,\,\,\,
,
\,\,\,\,\,\,\, a = \pm 1 
\label{kinksolphi4}
\EN 
where we have chosen the integration constant $x_0=0$. Using eq.\,(\ref{finalformulamass}), the classical mass of 
these topological configurations is given by 
\EQ
M_0\,=\,\int_{-\infty}^{\infty} \epsilon(x) \,dx \,=\,\frac{\sqrt{2}}{3} \,
\frac{m^3}{\lambda^2}  
\,\,\,.
\EN 
Keeping into account the finite quantum correction of this quantity \citep{DHN,raj}, the kink mass becomes
\EQ
M \,=\,\frac{ \sqrt{2}}{3} \,\frac{m^3}{\lambda^2} - m 
\left(\frac{3}{\pi \sqrt{2}} - \frac{1}{2 \sqrt{6}}\right) 
+ {\cal O}(\lambda) \,\,\,.
\label{mass1phi4}
\EN 
Conveniently defining  
\[
c =  \left(\frac{3}{2\pi} - \frac{1}{4 \sqrt{3}}\right) > 0 \,\,\,,
\]
and the dimensionless quantities
\EQ
g = \frac{3 \lambda^2}{\pi m^2}
\,\,\,\,\,\,\,\,
;
\,\,\,\,\,\,\,\,
\xi \,=\,\frac{g}{1 - \pi c g} \,\,\,, 
\label{definitiong}
\EN 
the mass of the kink can be finally expressed as 
\EQ
M \,=\,\frac{\sqrt{2} m}{\pi \,\xi}\,\,\,.
\label{newmassphi4}
\EN
In this example the kink and the anti-kink solutions are equal functions (up to a sign), and therefore their Fourier transforms have the same poles. Hence, the spectrum of the neutral particles will be the same on both vacua $\mid \pm 1 \,\rangle$, in agreement with the $Z_2$ symmetry of the model, and is given by 
\begin{equation}
m^{B}_{\pm, n} \,=\, 2 M\,\sin\left[n\,\frac{\pi \xi}{2}\,\right]
\hspace{5mm}, \hspace{5mm} n=1,2,\ldots N^B_{\pm} \,\,\,. 
\label{massphi4}
\end{equation}
The interested reader can find a more detailed analysis of this bosonic sector in \citep{GMneutral}. 

\vspace{1mm}
Let's now add the fermion and compute its spectrum (for a previous semi-classical study of this problem, 
close to the spirit of the DHN approach \citep{DHN}, see \citep{Campbell}). Since $V(\varphi)$ -- evaluated on the kink solution -- must assume opposite values at $x \rightarrow \pm \infty$, the simplest way to satisfy this requirement consists of a potential simply proportional to the bosonic field $\varphi(x)$ itself
\EQ
V(\varphi) \,=\, g \,\varphi(x) \,\,\,, 
\EN 
where $g$ is another coupling constant. So, for the asymptotic values of $V$ at $x\rightarrow \pm \infty$ we have 
\EQ
v_{a} = - v_{-a} \,=\, g  \,\hat\varphi\,\,\,.
\EN
Let's now briefly analysed the fermionic spectrum in the $Q=0$ and $Q=1$ sectors. 

\vspace{3mm}

\noindent
{\bf Fermionic bound states in $Q=0$ sector}. Given $V(\varphi)$, we can compute the fermionic zero mode, which in 
this case can be expressed in a closed simple form  
\begin{figure}[b]
\vspace{10mm}
\psfig{figure=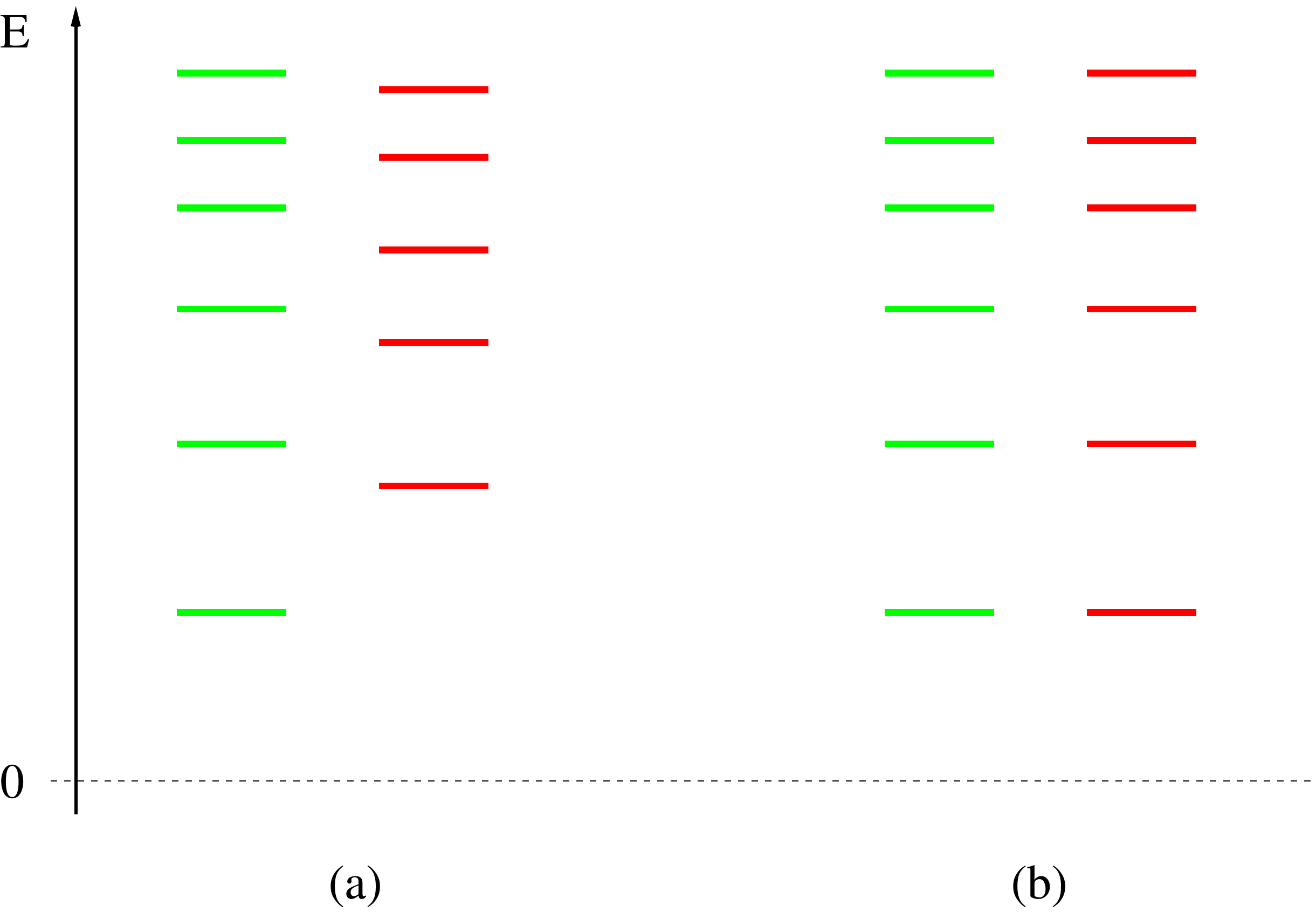,height=6cm,width=9cm}
\vspace{1mm}
\caption{{\em Spectrum of boson and fermion bound states in the $Q=0$ sector for: (a) generic values of the couplings $g$ and $\lambda$;  
(b) for $g = \lambda$}} 
\label{spectracomparing}
\end{figure}
\EQ
\psi_{-a,a}^{(0)}(x) \,= \, A\, \hat\psi \,\left(
\frac{1}{\cosh\frac{m x}{\sqrt{2}}}
\right)^r \,\,\,, 
\EN
where $r$ is the ratio of the couplings 
\EQ
r \,=\,\frac{g}{\lambda} \,\,\,.
\EN
Posed $\omega_a = \sqrt{2} m$, its behaviour at $x \rightarrow - \infty$ is given by 
\EQ
\psi_{a,a}^{(0)}(x) \simeq 2^{r}\, A \, \hat\psi \,e^{v_a x} \,\left[1 - r \,e^{{\omega_a x}} + r (r +1)  \,e^{2 \omega_a x} + \cdots \right] \,\,\,.
\EN
Hence the poles of the matrix element $G_{-a a}(\theta)$ are located at 
\EQ
\theta_n \,=\,i \pi \left(1 - \eta - n \,\xi\right) \,\,\,,
\EN 
with 
\EQ
\eta \,=\,\frac{v_a}{\pi M} 
\hspace{5mm}
,
\hspace{5mm}
\xi \,=\frac{\omega_a}{\pi M} \,\,\,, 
\EN
and the masses of the fermionic bound states at both vacua $\mid \pm \rangle$ are given by 
\EQ
m_{\pm,n}^F \,=\,2 M \,\sin\left(\frac{\pi(\eta + n \xi)}{2}\right) 
\hspace{5mm}, \hspace{5mm} n = 1, 2, \ldots, N^F_{\pm} \,\,\,. 
\label{spectrofermion}
\EN 
Comparing this spectrum with the one of bosons, eq.\,(\ref{massphi4}), we see that the masses of bosons and fermions, as well as their number, 
are in general different, see Figure \ref{spectracomparing}. There is however a special case where the two spectra match exactly: it is when $\eta = \xi$, alias when $g = \lambda$. As we will see in Section \ref{SUSY}, this is not a coincidence but it turns out to be the condition under which the theory is invariant under a supersymmetric transformation. 

\vspace{3mm}

\begin{figure}[t]
\psfig{figure=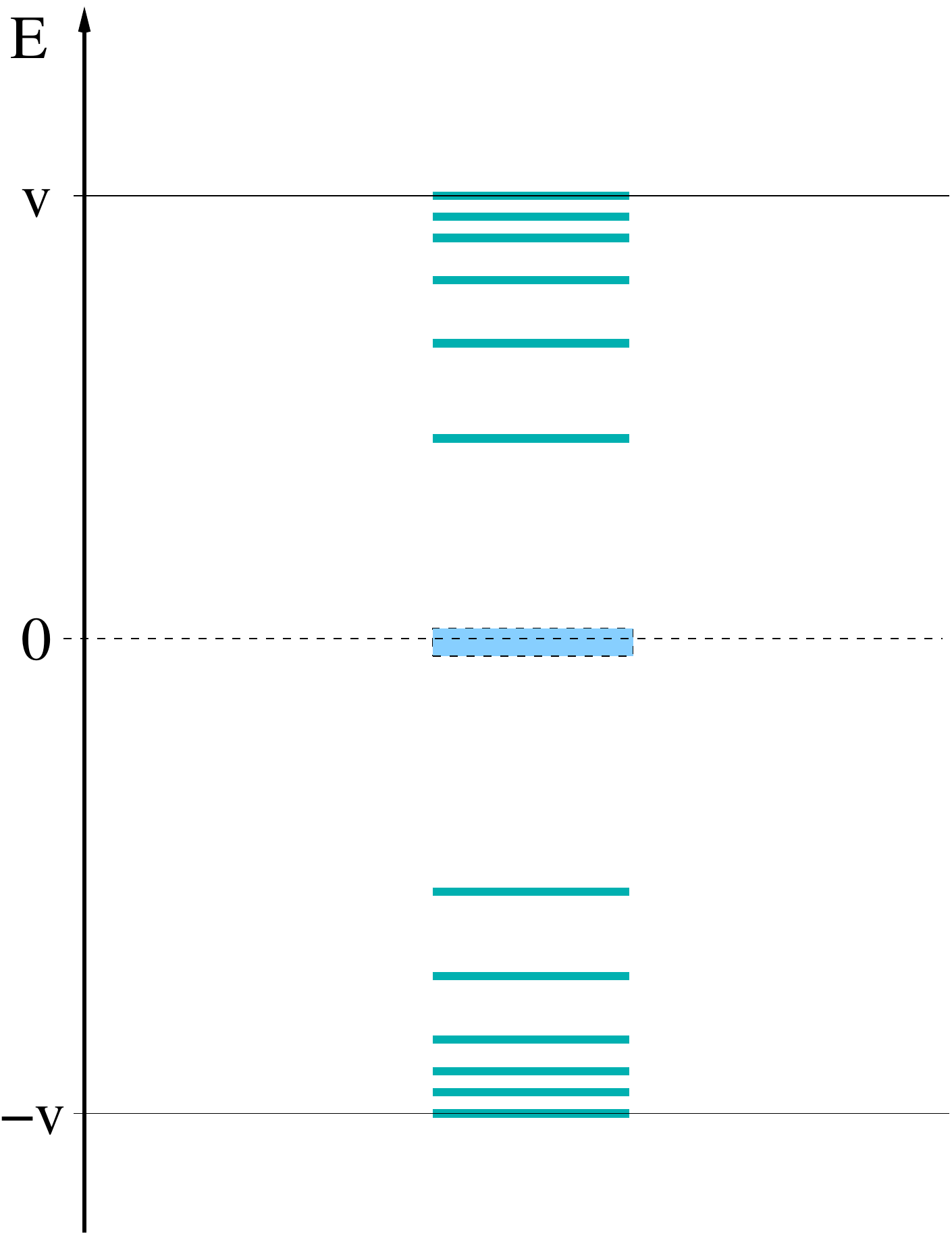,height=6cm,width=6cm}
\vspace{1mm}
\caption{{\em Discrete spectrum of fermions in the $Q=1$ sector, including the zero mode for $g/\lambda = 6$}} 
\label{discretefermionspectrum}
\end{figure}

\noindent
{\bf Fermionic bound states in $Q= \pm 1$ sector}. In the case of the symmetric wells it is possible to solve in a close form the 
set of Schr\"{o}dinger equations (\ref{SchroSUSY}) and find the fermion spectrum in the first non-trivial $Q=\pm 1$ topological sectors. 
The computation can be found in the original paper \citep{DHN} or in a more recent one \citep{iranian}: in the latter, in particular,  to get the spectrum 
of the Dirac Hamiltonian it has been used the technique of shape-invariant potentials. Here we simply report the final results for the bound state energies which depends upon only on the ratio $r$ of the two coupling constants 
\EQ
E_k \,=\,\pm \frac{m}{\sqrt{2}}\, \sqrt{2 r k - k^2} 
\,\hspace{3mm}
,
\hspace{3mm}
k = 1,2,\ldots \left[ r \right]
\EN 
The upper value of these energies coincides with $E_{max} = v_a$, which is the effective mass of the fermion, so the entire discrete spectrum is in the interval $(-E_{max},E_{max})$ as shown in Figure \ref{discretefermionspectrum}. Notice that such bound states only exist when $ r > 1$, otherwise they are absent. In particular, there is none of them at the supersymmetric point $g \,=\,\lambda$. 

\section{Asymmetric wells}
\label{asym}
\noindent
It is interesting to see what happens when the bosonic potential has two asymmetric wells $\mid {\bf a} \rangle$ and $\mid {\bf b} \rangle$: 
in this case the asymptotic behaviors of the kink at $x \pm \infty$  are different and therefore one should expect to find two different spectra
piling up on the two vacua. For the purely bosonic model this case has been discussed in \citep{GMneutral}, where it has been shown 
in particular that there could exist a range of the coupling constant where one vacuum (say $\mid {\bf a} \rangle$ ) has no bound states on top of it, 
while the other vacuum $\mid {\bf b} \rangle$ can have instead one or more. Such a situation is realised, for instance, in the 
Tricritical Ising Model, once perturbed away from criticality by the sub-leading magnetic field \citep{LMC,CKM}. Here we analyse 
such an asymmetric potential when there are also fermions. Let's first briefly remind the phenomenology of the purely bosonic case. 

\begin{figure}[b]
\psfig{figure=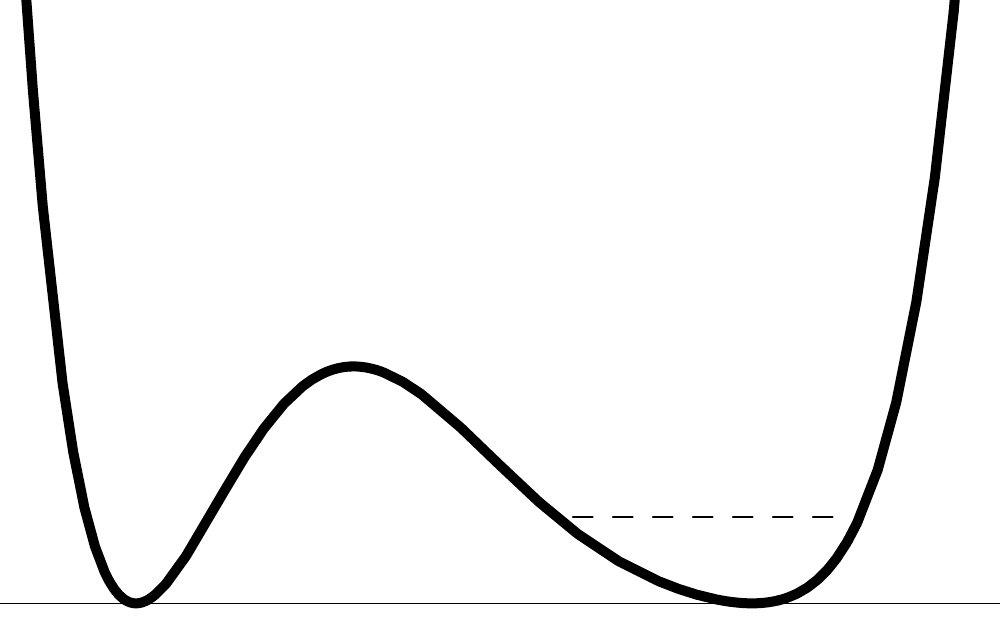,height=3cm,width=4cm}
\vspace{1mm}
\caption{{\em Example of $\varphi^6$ potential with two asymmetric wells and a bound state only on one of them.}}
\label{potential6}
\end{figure}

\vspace{3mm}
As bosonic potential $U(\varphi)$ of this model we choose 
\EQ
U(\varphi) \,=\,\frac{\lambda^2}{2} \, \left(\varphi+a\frac{m}{\lambda}\right)^2 \, \left(\varphi - b\frac{m}{\lambda}\right)^2\, 
\left(\varphi^2 + c\frac{m}{\lambda}\right) \,\,\,.
\label{Uphi6}
\EN
Rescaling the field as $\varphi(x) \rightarrow \frac{\lambda}{m} \varphi(x)$, we can bring the Lagrangian in the form 
\EQ
{\cal L} \,=\,\frac{m^6}{\lambda^4} \left[\frac{1}{2} (\partial \varphi)^2 - 
\frac{1}{2} (\varphi+a)^2 (\varphi-b)^2 
(\varphi^2 + c) \right]\,\,\,.
\label{newphi6}
\EN 
The minima of $U(\varphi)$ are localised at $\varphi_0^{(0)} = - a$ and $\varphi_1^{(0)} = b$ and the corresponding ground states will be denoted by $\mid {\bf a} \rangle$ and $\mid {\bf b} \rangle$. The curvature of the potential at these points is given by 
\EQ
\begin{array}{lll}
U''(-a) & \equiv & \omega^2_0 = (a+b)^2 (a^2 + c) \,\,\,;\\
U''(b) & \equiv & \omega^2_1 = (a+b)^2 (b^2 + c)\,\,\,, 
\end{array}
\label{curvature}
\EN 
and therefore for $ a \neq b$, we have two asymmetric wells, as shown in Figure \ref{potential6}. From now on we choose the curvature at the vacuum $\mid {\bf a} \rangle$ to be higher that the one at the vacuum $\mid {\bf b} \rangle$, i.e. a condition realised when $a > b$. 

\vspace{1mm}

The kink equation is given in this case by 
\EQ
\frac{d\varphi}{d x} \,=\,\pm (\varphi + a) (\varphi - b) \,\sqrt{\varphi^2 + c}\,\,\,.
\label{kinkphi6}
\EN 
The kink interpolates between the values $-a$ (at $ x = -\infty$) and $b$ (at $x = +\infty$) while the anti-kink does the viceversa. However, the behaviour of the anti-kink at $x = -\infty$ is different from the one of the kink, since the two vacua are approached differently: indeed, 
$\varphi_{ab}(x)$ approaches the vacuum $\mid {\bf a} \rangle$ as 
\EQ
\varphi_{ab}(x) \simeq -a + \hat A \exp(\omega_a x) 
\,\,\,\,\,\,\,
,
\,\,\,\,\,\,\,
x \rightarrow - \infty 
\EN 
while the anti-kink approaches instead the vacuum $\mid {\bf b} \rangle$ as 
\EQ
\varphi_{ba}(x) \simeq b - \hat B \exp(\omega_b x) 
\,\,\,\,\,\,\,\,
,
\,\,\,\,\,\,\,\,
x \rightarrow - \infty
\EN 
where $\hat A$ and $\hat B$ are two positive constants. Since $\omega_0 \neq \omega_1$, the asymptotic behaviour of the two solutions 
ends up in the following poles in their Fourier transform 
\begin{eqnarray}
{\cal F}(\varphi_{ab}) & \rightarrow & \frac{\hat A}{\omega_a + i k} \nonumber \\
& & \label{polephi6}\\
{\cal F}(\varphi_{ba}) & \rightarrow & \frac{- \hat B}{\omega_b + i k} \nonumber 
\end{eqnarray}
A dimensional analysis fixes the mass of the kink to be  
\EQ
M\,=\,\frac{m^5}{\lambda^4} \,\alpha \,\,\,,
\EN 
where $\alpha$ (that depends upon  $a, b, c$) can be computed taking the integral (\ref{finalformulamass}). Using this expression of the mass and   
expressing the poles given above in the variable $\theta$ we have 
\begin{eqnarray}
\theta^{(a)} \,& \simeq & \,i\pi \left( 1 - \omega_a \,\frac{m}{\pi M}\right) = i\pi 
\left(1 - \omega_a \,\frac{\lambda^2}{\alpha m^4}\right)\,\,\,,
\nonumber \\
& & \\
\theta^{(b)} \,& \simeq & \,i\pi \left( 1- \omega_b \,\frac{m}{\pi M}\right) 
\,=\, i\pi \left(1 - \omega_b \,\frac{\lambda^2}{\alpha m^4}\right) \,\,\,. \nonumber 
\end{eqnarray}
Therefore, if we choose the coupling constant in the range 
\EQ
\frac{\pi\alpha}{\omega_a} < \frac{\lambda^4}{m^4} < \frac{\pi \alpha}{\omega_b}    \,\,\,,
\label{range}
\EN 
the first pole will be out of the physical sheet whereas the second will still remain inside it. Namely, the theory will have only one neutral bosonic 
bound state, localised at the vacuum $\mid {\bf b} \,\rangle$. Said differently, an antikink-kink configuration produces a bound state whereas a kink-antikink does not. 

\begin{figure}[t]
\psfig{figure=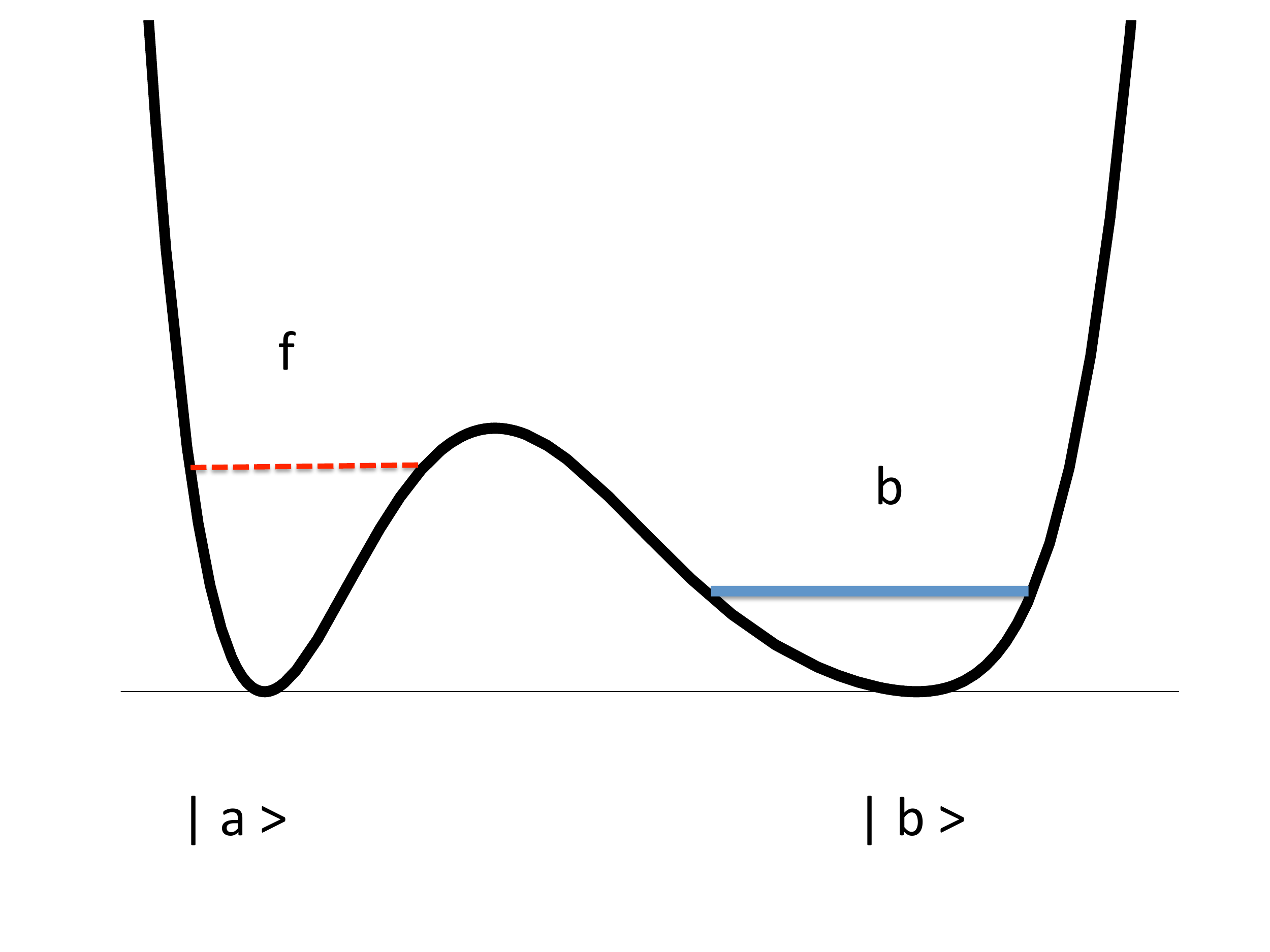,height=4cm,width=5cm}
\vspace{1mm}
\caption{{\em Asymmetric potential, with one fermion bound state at the vacuum $\mid {\bf a} \rangle$ and no boson bound state, and 
a boson bound state at the vacuum $\mid {\bf b} \rangle$ but no fermion one.}}
\label{reversefig}
\end{figure}

\vspace{1mm}
Let's now consider how the fermion may change the previous picture. As potential term $V(\varphi)$ we consider once again for simplicity  
$ V(\varphi) \,=\, g \varphi(x)$, so that 
\EQ 
\begin{array}{lll}
v_a \,\equiv \,-V(-a) &\,=\, & g \,a\, \frac{m}{\lambda}   \,\,\,, \\
\\
v_b \,\equiv\, \,\,\,V(b) &\,=\,& g \,b \frac{m}{\lambda}\,\,\,.
\end{array}
\EN
This time there will be a zero-mode $\psi_{ab}^{(0)}$ corresponding to the kink configuration $\varphi_{ab}(x)$ and another one $\psi_{ba}^{(0)}$ relative to the antikink $\varphi_{ba}(x)$, with the behavior at $x\rightarrow \mp\infty$ of one equal to that one of the other at $x \rightarrow \pm \infty$, so that 
\begin{eqnarray}
\psi_{ab}^{(0)}(x) & \,=\, & A \,\hat\psi\, \, \exp\left[- \,\int_{x_0}^x V_{ab}(t) \, dt \right] \,\rightarrow \, 
A \,\hat\psi\, \, e^{v_a x} \, \left(1 + \sum_{n=1} t_{n}\, e^{n \omega_a x}\right)  
\,\,\,\,\,,\,\,\,\, x\ \rightarrow - \infty \\
\psi_{ba}^{(0)}(x) & \,=\, & A \,\hat\psi\, \, \exp\left[- \,\int_{x_0}^x V_{ba}(t) \, dt \right] \,\rightarrow\, 
A \,\hat\psi\, \, e^{v_b x} \, \left(1 + \sum_{n=1} t_{n}\, e^{n \omega_b x}\right)  
\,\,\,\,\,,\,\,\,\, x \rightarrow - \infty 
\label{diffzeroodeexpterm}
\end{eqnarray}
Therefore, posing $\eta_k = v_k/(\pi M)$ and $\xi_k = \omega_k/(\pi M)$ ($k = a,b$), we find two different fermion spectra on the two vacua 
\EQ
m_{k}^F(n_k) \,=\,  2 M \sin\left(\frac{\pi}{2} (\eta_k + n_k \xi_k)\right) \,\,\,\,\,\,\,,\,\,\,\,\,\, n_k =1,2, \ldots \left[\frac{1 - \eta_k}{\xi_k} \right]
\,\,\,.
\EN
Varying the coupling constants $\lambda$ and $g$ one can get various and also curious situations at the two vacua: for instance, imagine that we are in the condition (\ref{range}), where there is no bosonic bound state on the vacuum $\mid {\bf a} \rangle$ and only one on the vacuum $\mid {\bf b} \rangle$. 
But we can easily revert the situation for what concerns the fermionic spectrum! Alias, if we choose the ratio of the coupling constants $g$ and $\lambda$ such that it satisfies the condition (where ${\cal C} = \pi \alpha m^4/\lambda^4$) 
\EQ
\frac{\cal C}{b} \, < \frac{g}{\lambda} < \frac{\cal C}{a} \,\,\,,
\EN 
then the vacuum $\mid {\bf a} \rangle$ will have a fermionic bound state while the vacuum $\mid {\bf b} \rangle$ has none, as shown in Figure \ref{reversefig}. 

\section{Supersymmetric theory}\label{SUSY}
\noindent
Let's consider now QFT's which are invariant under a $N=1$ supersymmetry transformation. In this case, the bosonic and 
fermionic fields, together with a real auxiliar field $F(x)$ can be conveniently organised into a real  superfield  
 $\Phi(x,\theta)$ that admits the expansion 
\EQ
\Phi(x,\theta) \,=\,\varphi(x) + \bar\theta\,\psi(x) + \frac{1}{2} \bar\theta\,\theta F(x) \,\,\,.
\EN 
The space coordinates $x^{\mu}=(x^0,x^1)$ and the two real Grassmann coordinates 
$\theta_{\alpha} = (\theta_1,\theta_2)$ describe the $N=1$ superspace. Under a translation in superspace 
 \EQ
 x^{\mu} \rightarrow x^{\mu} + i \bar\xi \gamma^{\mu} \theta 
\,\,\,\,\,\,\,\,
,
\,\,\,\,\,\,\,\,
\theta_{\alpha} \rightarrow \theta_{\alpha} + \xi_{\alpha} 
\label{translationsusy}
\EN 
the variation of the superfield is given by 
\EQ
\delta \Phi(x,\theta)\,=\,\bar \xi_{\alpha} \,{\cal Q}_{\alpha}\,\Phi(x,\theta) \,\,\,,
\label{susytran}
\EN 
with ${\cal Q}_{\alpha} = \partial/\partial \bar\theta_{\alpha} + i (\gamma^{\mu} \theta)_{\alpha}\,
\partial_{\mu}$. The most general action invariant under the supersymmetric transformation 
(\ref{susytran}) is given by 
\EQ
{\cal A} \,=\,\int d^2 x\, d^2 \theta \left[
\frac{1}{4} (\bar D_{\alpha} \Phi) \,D_{\alpha} \Phi + W(\Phi) \right] \,\,\,,
\label{susyaction}
\EN 
where $\int d^2\theta \bar \theta \theta = 2$, with the covariant derivative $D_{\alpha}$ 
given by 
\EQ
D_{\alpha} \equiv \frac{\partial}{\partial \bar\theta_{\alpha}} - 
(i \partial_{\mu} \gamma^{\mu} \theta)_ {\alpha} \,\,\,.
\EN 
$W(\Phi)$ is the so-called superpotential, that we assume to be an analytic function of $\Phi$. 
Integrating on the Grassmann variables, one arrives to the following expression of the action 
\EQ
{\cal A}\,=\,\int d^2 x \left\{
\frac{1}{2} \left[ (\partial_{\mu} \varphi)^2 + i \bar\psi \gamma^{\mu} \partial_{\mu}
\psi + F^2 \right] + F\,W^{'}(\varphi) - \frac{1}{2} W^{"}(\varphi) \bar\psi\psi \right\}\,\,\,,
\EN 
where $W^{'}(\varphi) = dW(\varphi)/d\varphi$, etc.  Finally, elimitating the auxiliary field $F$ from 
its algebraic equation of motion, i.e. substituting $F \rightarrow - W^{'}(\varphi)$ in the above 
expression, and rescaling for convenience the fermion field as $\psi \rightarrow \sqrt{2} \psi$, 
it yields the general form of the lagrangian density for a supersymmetric theory given by
\EQ
{\cal L} \,=\,
\frac{1}{2} \left[ (\partial_{\mu} \varphi)^2 - [W^{'}(\varphi)]^2\right] + 
+ i \bar\psi \gamma^{\mu} \partial_{\mu}
\psi  - \frac{1}{2} W^{"}(\varphi) \bar \psi \psi \,\,\,.
\label{finalsusy}
\EN
Associated to the transformation (\ref{translationsusy}) there is the conserved supercurrent 
\EQ
J^{\mu}_{\alpha}(x) \,=\,(\partial_{\nu} \varphi) (\gamma^{\nu} \gamma^{\mu} \psi)_{\alpha} - i F 
(\gamma^{\mu} \psi)_{\alpha} \,\,\,, 
\label{supercurrent}
\EN 
and the conserved supercharges 
\EQ
Q_{\alpha} \,=\,\int dx^1\,J_{\alpha}^0 \,\,\,.
\label{supercharge}
\EN 
Together with the stress-energy tensor 
\EQ
{\cal T}^{\mu\nu}(x) \,=\,i\,\bar\psi \,\gamma^{\mu} \partial^{\nu} \psi + 
\partial^{\mu} \varphi \,\partial^{\nu} \varphi - \frac{1}{2} g^{\mu\nu} \left[
(\partial_{\alpha} \varphi)^2 - F^2\right] 
\,\,\,,
\label{stressenergy}
\EN 
and the topological current 
\EQ
\xi^{\mu}(x) \,=\,-\epsilon^{\mu\nu} F\,\partial_{\nu} \varphi \,=\,\epsilon^{\mu\nu} \partial_{\nu} W(\varphi) 
\,\,\,.
\EN 
they close the supersymmetry algebra \citep{OliveWitten}
\EQ
\{Q_{\alpha},\bar Q_{\beta}\} \,=\,2 (\gamma_{\lambda})_{\alpha\beta}\,P^{\lambda} + 
2 i (\gamma_5)_{\alpha\beta}\,{\cal Z} \,\,\,,
\label{susyalgebra}
\EN 
where $P^{\lambda} = \int {\cal T}^{0 \lambda}(x) \,dx^1$ is the total energy and momentum, 
whereas ${\cal Z}$ plays now the role of the topological charge 
\EQ
{\cal Z}_{ab} \,=\,\int \xi^0(x)\, dx^1 \,=\,\left[W(\varphi)\right]^{+\infty}_{-\infty} \,\equiv\,
W(\varphi_b) - W(\varphi_a)  \,\,\,.
\label{topologicalcharge}
\EN 
Although closely related to the topological charge introduced in (\ref{topologicalcharge0}), depending on the model ${\cal Z}$ may however differs from it. We will comment on this fact on the models considered later. A convenient and explicit form way of the supersymmetry algebra (\ref{susyalgebra}) is the following 
\EQ
Q_+^2\,=\,P_+ 
\,\,\,\,\,\,\,
,
\,\,\,\,\,\,\,
Q_-^2 \,=\,P_- 
\,\,\,\,\,\,\,
,
\,\,\,\,\,\,\,
Q_+ \, Q_- + Q_- \, Q_+ \,=\, 2 {\cal Z} \,\,\,,
\label{susyconvenient}
\EN 
where we have used the notation $(Q_1,Q_2) \equiv (Q_-,Q_+)$ and $P_{\pm} = P_0 \pm P_1$. 

\section{General results in SUSY theories} 
\label{generalSUSY}
\noindent
The existence of an algebra that relates the bosonic and fermionic sectors poses of course important constraints on the theory. Here we briefly collect some of these consequences. First of all, comparing the Lagrangian eq.\,(\ref{finalsusy}) with the original one (\ref{Lagrangianfermion}), 
we see that in $N=1$ SUSY model both potentials $U(\varphi)$ and $V(\varphi)$ come from the same function $W(\varphi)$, alias 
\EQ
\begin{array}{l}
U(\varphi) \,=\, \frac{1}{2} (W'(\varphi))^2 \,\,\,,\\
\\
V(\varphi) \,=\, W''(\varphi) \,\,\,.
\end{array}
\EN 
$U(\varphi)$ is then intrinsicaly positive and it is well known its relation with the spontaneously breaking of supersymmetry \citep{Witten}: in fact,  when $U(\varphi)$ has zeros, they are the true vacua of the theory and supersymmetry is unbroken; viceversa, if $U(\varphi)$ has local minima that are not zeros of this function, supersymmetry will be spontaneously broken at these minima. The local minima may be regarded as meta-stable vacua, as far as it exists a true vacuum somewhere in the landscape of $U(\varphi)$. 

\vspace{1mm}
Focusing now the attention on the bosonic and fermionic bound states in $Q = 0$ sector, let's show that the general pattern of their spectrum 
is fixed by the following two results. 
\begin{enumerate}
\item {\bf An identity concerning the potentials}. For all SUSY models it is easy to prove that for the fermionic and bosonic potentials we have   
\EQ
v_a \,=\,\pm \omega_a \,\,\,. 
\label{identity}
\EN 
Such an identity is easy to prove since 
\EQ
v_a \,=\,V(a) \,=\,W''(a) \,\,\,,
\EN 
while 
\EQ
\omega_a^2 \,=\, \frac{d^2 U}{d\varphi^2} \,=\,(W''(a))^2 +  W'(a) \,W'''(a) \,=\, (W''(a))^2 \,\,\,,
\EN
where the last step comes from $W'(a) =0$, since $\varphi_a$ is one of the vacua of $W'(\varphi)$.  
\item {\bf Exact value of the classical mass of kinks}.
In SUSY theories the classical mass of the kinks can be exactly computed entirely in terms of the algebra \citep{OliveWitten}. 
Indeed the minima of $U(\varphi)$ are connected by supersymmetric kinks, solutions of the so-called BPS equations 
\EQ
(Q_+ \pm Q_-) \,\mid\,K_{ab} \,\rangle \,=\,0 \,\,\,,
\label{killingBPS}
\EN 
where the $\pm$ refers to the kink $\mid K_{ab}\rangle$ and the antikink $\mid K_{ba}\rangle$ respectively. 
The bosonic component of these equations is just the familiar condition (\ref{kinkequation})  
\EQ
\frac{d\varphi_{ab}}{dx} \,=\,\pm\,F(\varphi_{ab}) \,=\,\pm\,W^{'}(\varphi_{ab}) \,\,\,. 
\label{BPSbound}
\EN 
Thanks to the identity 
\EQ
P_+  + P_- \,=\, (Q_+ \pm Q_-)^2 \mp 2 {\cal Z}_{ab} \,\,\,
\label{remarkableidentity}
\EN 
that follows from the supersymmetry algebra (\ref{susyconvenient}), the classical mass $M_{ab}$ of the kinks is then simply expressed 
by their topological charge:  for the kink (anitkink) at rest, the left hand side is given by $P_+ + P_- = 2 M$, whereas the right hand side, using eq.\,(\ref{killingBPS}), is equal to $\mp 2 Z_{ab}$, therefore \citep{OliveWitten} 
\EQ
M_{ab} \,=\, |{\cal Z}_{ab}| \,\,\,.  
\label{masskink}
\EN 
Notice that this expression is a particular case of the general formula (\ref{finalformulamass}), given that in supersymmetry $\sqrt{2 U(\varphi)}$ is equal to $W(\varphi)$ and the topological charge ${\cal Z}_{ab}$ is expressed by eq.\,(\ref{topologicalcharge}). Let's mention that it has been long disputed whether this result -- true at the classical level -- remains still valid once included quantum corrections (the interested reader may find a vivid summary of the story of this problem and the relevant references in the Chapter 3 of the book \citep{Shifman}). The final word on this problem seems to be as follows: at the quantum level, the mass of the kinks gets corrected $M_{ab} \rightarrow M_{ab} + \delta M_{ab}$ but nevertheless satisfies an equation as (\ref{masskink})) if one takes into a proper account that the superpotential $W$ (and then $Z$) gets a correction too. 
\end{enumerate}

\vspace{3mm}
\noindent
{\bf Degeneracy of the bosonic and fermionic bound states}. We can now used the two results given above to reach a very general conclusion 
for the spectra of supersymmetry theories. Namely, given the identity $\omega_a = \mid v_a \mid$ 
and the expression (\ref{masskink}) of the mass of the kinks, for SUSY theories we have  
\EQ
\eta_a \,=\,\xi_a\,\,\,,
\EN 
and therefore the spectra (\ref{universalmassformula}) and (\ref{numberfermionicboundstate}) of bosonic and fermionic bound states at each 
true ground state $\mid {\bf a} \rangle$ of zero energy are always necessarily paired.

\section{Integrable SUSY Models}
\label{integrableSUSY}
\noindent
There is a set ot theories where we can test our results: those are the integrable exactly solvable model. Here we focus the attention on a paradigmatic example of these theories, alias the SUSY Sine-Gordon model, whose superpotential $W(\Phi)$  
\EQ
W(\Phi) \,=\, \frac{m}{\lambda^2} \,\cos(\lambda \Phi) \,\,\,
\EN 
gives rise to the SUSY Sine-Gordon Lagrangian 
\EQ
{\cal L} \,=\,\frac{1}{2} (\partial_\mu\varphi)^2 + \frac{m^2}{2 \lambda^2} \,\sin^2\lambda\varphi + i \bar\psi\gamma^\mu\partial_\mu\psi -  
m \bar\psi \psi \cos\lambda \varphi \,\,\,.
\label{SUSYSG}
\EN
A big deal is known about this model, in many different contexts (see, for instance, \citep{sengupta,AhnSSG,tsvelik,bdptw,sakagami}): in particular, 
the issue of the finite correction to the kink mass (\ref{masskink}) has been clarified in \citep{BPS} while the exact $S$-matrix and the related 
spectrum have been determined in \citep{AhnSSG}. So, the exact spectrum of the $Q=0$ sector of this model is given by 
\EQ
m^B(n) \,=\,m^F(n) \,=\, 2 M \sin\left(\frac{n \hat\lambda^2}{2}\right) \,\hspace{5mm} , \hspace{5mm}\, n=1,2,\ldots 
\left[\frac{1}{\hat\lambda^2}\right]\,,
\label{exactspectrumSSGordon}
\EN 
where $M$ is the mass of the kink while $\hat\lambda^2$, the so-called renormalized coupling constant, is given by 
\EQ
\hat\lambda^2 \,=\, \frac{\lambda^2}{2\pi} \,\frac{1}{1-\frac{\lambda^2}{4\pi}} \,\,\,.
\label{renormalizationlambda}
\EN
Let's see how the spectrum (\ref{exactspectrumSSGordon}) is recovered by our semi-classical approach and, in the way to show this, 
also learning about other properties of this model which will be useful in the next section when we study the breaking of its integrability. 

\begin{figure}[t]
\psfig{figure=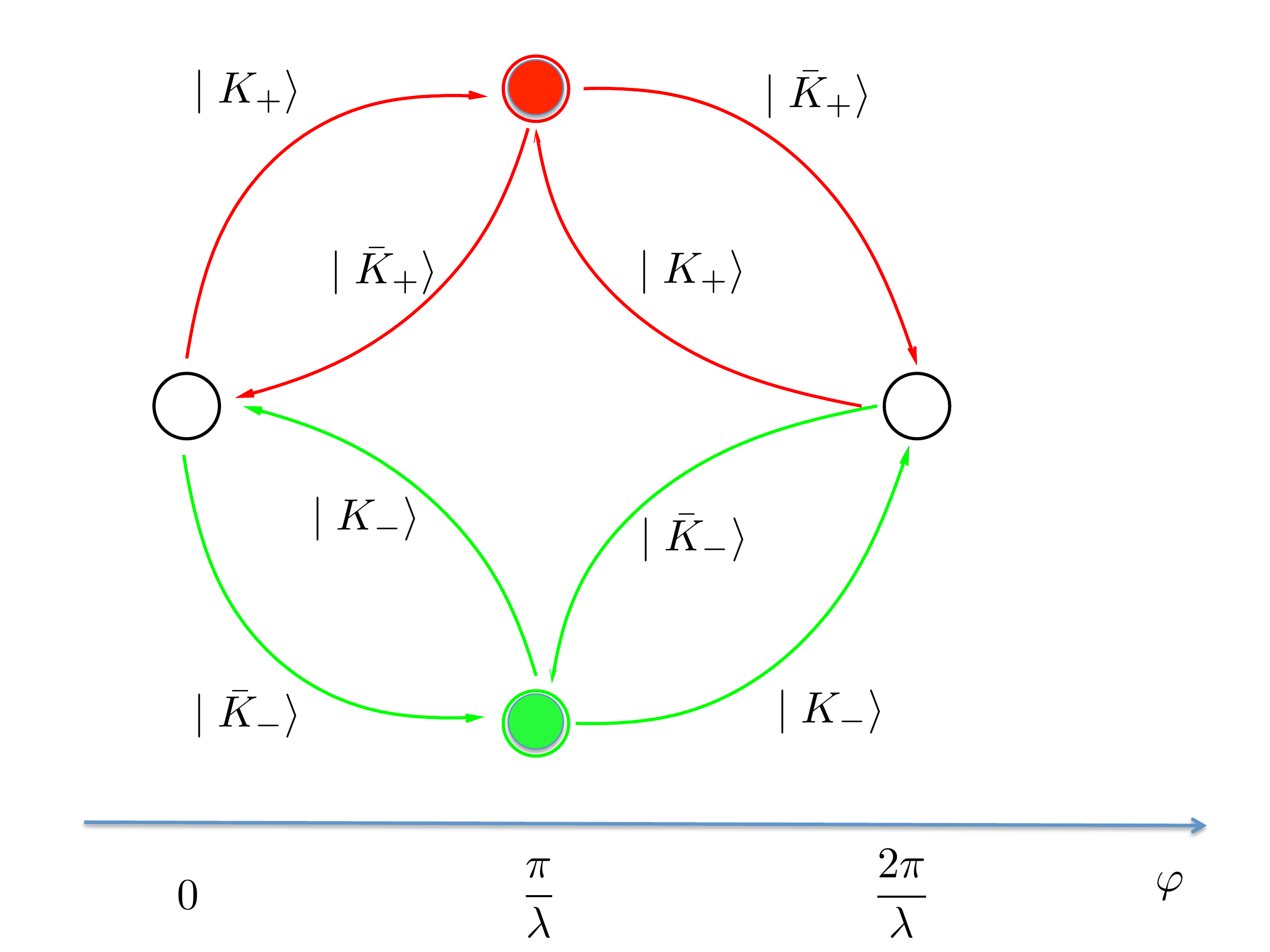,height=6cm,width=8cm}
\vspace{1mm}
\caption{{\em Structure and adjacency diagram of the vacua in the SUSY Sine-Gordon in the periodic interval of the model. This structure has to be periodically repeated for the other vacua.}} 
\label{SUSYSGVacua}
\end{figure}
The minima of the scalar potential $U(\varphi)$ are localised at $\varphi_k^{(0)}\,=\,k\,\pi/\lambda$. Since they are also the zeros of $ U(\varphi)$, they are supersymmetric vacua of the quantum theory. The theory (\ref{SUSYSG}) is invariant under the $Z_2$ parity $\varphi \rightarrow - \varphi$ and under the shift 
\EQ
\varphi \rightarrow \varphi + \frac{2\pi \,n}{\lambda} \,\,\,, 
\label{firstperiod}
\EN 
so we restrict our attention to the interval $(0, 2 \pi/\lambda$) of the field $\varphi$. The elementary kink and the anti-kink solutions of (\ref{kinkequation}) of the SSG model are explicitly given by 
\EQ
\varphi^{cl}_{\pm}(x) \,=\,\frac{2}{\lambda} \,\arctan (e^{\pm m x}) \,\,\,, 
\label{classicalsusykinks}
\EN 
and connect the adjacent vacua $\varphi_0^{(0)} = 0$ and $\varphi_1^{(0)} = \pi/\lambda$, all other kinks or antikinks of the model being equivalent to them. In this model ${\cal Z}$ differs from the usual topological charge $Q$, since the latter counts the kink number while the former counts the kink number only modulo two \citep{OliveWitten}. 

In addition to the explicit expression of the kinks, there is also a close formula also the fermionic zero mode, since 
the integral of $V(\varphi_{cl}(x)$ can be dealt as 
\EQ
\int_{x_0}^x V(\varphi^{cl}(x) dx \,=\,  \int_{\varphi^{cl}(x_0)}^{\varphi^{cl}(x)} V(\varphi)\, \left(\frac{d\varphi}{dx}\right)^{-1} \, dx  \,=\,  \int_{\varphi^{cl}(x_0)}^{\varphi^{cl}(x)} \frac{\cos t}{\sin t} \, dt \,=\, \log \left[\sin(\varphi^{cl}(x))\right] \,\,\,,
\EN
where we have used eq.(\ref{BPSbound}) for the derivative of the kink solution and choose $x_0$ such that $\varphi_{cl}(x_0) = \pi/(2 \lambda)$. 
In this way, the zero mode can be exactly expressed as 
\EQ
\psi_{01}^{(0)}(x) \,=\, A \, \hat\psi_0 \sin\left[2 \arctan( e^{m x})\right] \,=\,A \, \hat\psi_0 \,\frac{1}{\cosh ( m x)} \,\,\,, 
\EN
and, as a matter of fact, is equal to the zero-mode of the double well potential at the supersymmetric point $g \,=\,\lambda$. 

\vspace{1mm}
About the nature of the SUSY vacua, we can take every vacuum $\varphi_{2n}^{(0)} \,=\, 2 n \,\pi/\lambda$ to be non-degenerate
(these are the vacua where $V(\varphi_{2n}^{(0)}) > 0$), while we can unfold those located at $\varphi_{2n+1}^{(0)} \,=\,(2 n +1)\,\pi/\lambda$ 
(where $V(\varphi_{2n+1}^{(0)}) < 0$): the adjacency properties of the new vacua are shown in Figure \ref{SUSYSGVacua}. This way of unfolding the vacua is similar to what has been done in the Supersymmetry deformation of the Tricritical Ising Model \citep{Zamfrac}. The classical mass of these kinks, according to the formula (\ref{masskink}), is given by 
\EQ
M_{cl}\,=\, \frac{2 m}{\lambda^2} \,\,\,. 
\EN 
In references \citep{BPS} it has been computed the finite quantum correction of this mass 
\EQ
M_{cl} \rightarrow M\,=\, 2 \,m \,\left[\frac{1}{\lambda^2} - \frac{1}{4\pi}\right]
\,\,\,,
\EN 
an expression that can be written as 
\EQ
M \,\equiv \frac{m}{\pi \hat\lambda^2} \,\,\,,
\EN 
where $\hat\lambda$ is the effective coupling constant of the SUSY Sine-Gordon model given in (\ref{renormalizationlambda}). 
This formula closes now the circle, since all the minima of $U(\varphi)$ has a curvature equal to $\omega_a = m$ and therefore for 
the semiclassical parameter $\eta_a$ and $\xi_a$ we have  
\EQ
\xi_a \,=\,\eta_a\,=\,\frac{m}{\pi M} \,=\,\hat\lambda^2 \,\,\,. 
\EN
Hence, for this exactly solvable model, the semi-classical spectra (\ref{universalmassformula}) and (\ref{spectrofermion}) precisely coincide with the exact ones (\ref{exactspectrumSSGordon}).

\section{Non-integrable SUSY Models}
\label{nonintegrableSUSY}
\noindent
The SUSY Sine-Gordon model has provided an important check of our formulas for the mass spectrum of bosons and fermions. However the actual 
usefulness of the semi-classical formulas is of course in the absence of an exact solution of the model. This is, for instance, the case of the double well theory analysed in Section \ref{symmetriccase}: even though when $g \,=\, \lambda$ this model can be elegantly put in a SUSY form, with the super-potential 
\EQ
W(\Phi) \,=\,\lambda \Phi^3 - \frac{m^2}{2 \lambda} \Phi \,\,\,,
\label{WZmodel}
\EN 
it remains nevertheless non-integrable and therefore the discussion done in Section \ref{symmetriccase} becomes particularly valuable. 

While the model defined by eq.\,(\ref{WZmodel}) has no parameter and it is simply non-integrable, here we are interested to study a class of non-integrable SUSY models defined by a Lagrangian ${\cal L}(\rho)$ depending on a parameter $\rho$ that varies on the interval $(0,1)$, such that the corresponding theory is integrable at both extrema, $\rho =0$ and $\rho=1$, but non-integrable otherwise. A representative of this type of models is 
the multi-frequency SUSY Sine-Gordon model \citep{GMSUSYKINK}, with super-potential given by 
\EQ
W(\Phi,\rho) \,=\,m \,\left(\frac{1-\rho}{\lambda^2} \, \cos(\lambda \Phi) + \,
\frac{\rho}{\beta^2} \, \cos(\beta \Phi) \right) \,\,\,. 
\EN
The associated bosonic potential is given by 
\EQ
U(\varphi,\rho) \,=\, \frac{m^2}{2} \,\left(\frac{1 - \rho}{\lambda} \, \cos(\lambda \Phi) +  \,
\frac{\rho}{\beta} \, \sin(\beta \Phi) \right)^2 \,\,\,.
\label{newpotential}
\EN 
We have chosen a proper parameterization in such a way that the curvature at the origin is always given by $m^2$. In the flow of $\rho$ from $0$ to $1$ 
one expects that the spectra at the different vacua move correspondingly and we are interested to see whether we can reach a certain control of this situation. 

\begin{figure}[t]
\psfig{figure=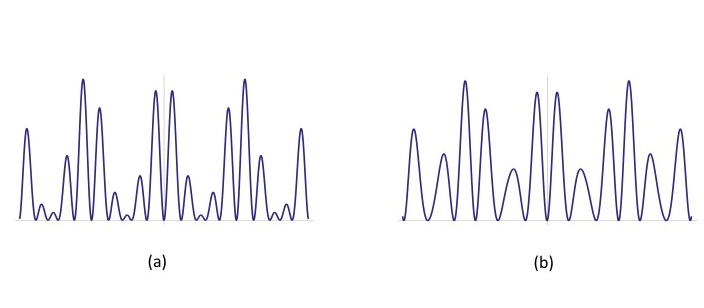,height=7cm,width=15cm}
\caption{{\em $V(\varphi,\rho)$ for an irrational value of $\omega$ (here the golden ratio) and different values of the deformation parameter $\rho$: (a) $\rho = 0.3$; (b) $\rho = 0.7$.}}
\label{noperiodicity}
\end{figure}

\vspace{1mm}
First of all it is obvious that a relevant parameter is the ratio of the two frequencies 
\EQ
\omega \,=\, \frac{\beta}{\lambda}\,\,\,,
\EN
since we can always rescale the bosonic field as $\varphi \rightarrow \varphi/\lambda$ and manage to have $\omega$ as frequency of the second term in (\ref{newpotential}). If $\omega$ is irrational, the potential (\ref{newpotential}) loses any periodicity in $\varphi$ and, apart the residue $Z_2$ summetry $\varphi \rightarrow - \varphi$, it generically gives rise to an infinite number of vacua that may move in a complicate way by varying the parameter $\rho$ (see, for instance, Figure \ref{noperiodicity}). Such a situation is the more complicated one but, using the continue fraction
representation of any real number in terms of rational numbers, it can be studied as a limit case of the rational ratio of the frequencies, a situation to which we turn our attention. 

Let's consider then the case when $\omega$ is a rational number, $\omega = \frac{p}{q}$ (say with $q > p$). In this case, the periodicity of the original potential stretches to 
\EQ
{\cal I} \,= \,\left(0,\frac{2\pi q}{\lambda}\right) \,\,\,.
\label{newperiodicity}
\EN 

\vspace{1mm}
As noticed in \citep{GMSUSYKINK}, an important characteristic of the field theory described by (\ref{newpotential}) is that the kinks initially present 
at $\rho =0$ also persist for finite values of $\rho$, {\em regardless} of the ratio  $\omega$ of the two frequencies -- a result that is strikingly different of what happens in the purely bosonic multi-frequency Sine-Gordon model \citep{DM}. The proof of this feature of the model, based on its Supersymmetry and pursued through the Form-Factor Perturbation Theory, can be found in \citep{GMSUSYKINK}. An intuitive way to understand this robustness of the kinks is in terms of the persistence of the original zeros of the bosonic potential $U(\varphi,\rho=0)$ for {\em finite values} of $\rho$. Moreover, as discussed below, when they disappear they do so in pairs and this happens only at some critical values $\rho_m$ of the parameter $\rho$. This observation helps in understanding the evolution of the kinks and the number of stable particles at any given $\rho$. To make it quantitative, call  ${\cal M}_0(\rho)$ the number of zeros of $U(\varphi,\rho)$ at a given value of $\rho$ in the extended interval of periodicity (\ref{newperiodicity}).  Clearly 
\EQ
{\cal M}_0(0) = 2 q +1
\hspace{5mm} ; 
\hspace{5mm}
{\cal M}_0(1) = 2 p +1
\,\,\,,
\EN
so that from the beginning to the end of the flow there has been a total variation $\Delta {\cal M}_0 = 2 (q - p)$ of the zeros. Since the existence of the kinks is linked to the zeros of $U(\varphi,\rho)$, a variation of the latter implies the disappearing of some of the former along the flow from $\rho = 0$ to $\rho =1$. Being the kinks topological excitations, their disappearance will signal the occurrence of a phase transition at some critical values of the parameter, $\rho = \rho_c^{(m)}$,  which are the values where the number of zeros jumps by a step of 2 units. In fact, two zeros collide and then move on imaginary values;  the barrier between these zeros vanishes and correspondingly the kink/antikink that connect the two colliding vacua become massless. 
    
The persistence of the kinks of the original theory at weak coupling does not mean that their mass remain untouched. In particular, 
the mass degeneracy of all the kinks of the integrable SUSY Sine-Gordon model at $\rho =0$ is split when $\rho$ is switched on, and we will have 
{\em long-kinks} (overpassing higher barriers) and {\em short kinks} (overpassing lower barriers), as shown for instance in Figure 9.  The actual computation of the new classical masses of the kinks can be done by employing their topological charges, i.e. by using eqs.\,(\ref{topologicalcharge}) and (\ref{masskink}), together with the positions of the new zeros of $U(\varphi,\mu)$.  

\vspace{1mm}
We can now write down the steps to take in order to see how the spectrum evolves by varying $\rho$: 
\begin{enumerate}
\item Determine in the interval (\ref{newperiodicity}) the zeros $\varphi_n^{(0)}(\rho)$ ($n=0,1,\ldots, k)$ of $V(\varphi,\rho)$ by moving $\rho$. 
Such zeros are associated to the stable vacua $\mid {\bf n} \rangle$ at that value of $\rho$. Their number depends on $\rho$ and changes discontinuously of $2$ units at some critical values $\rho_m^{(c)}$.  Notice that the zeros placed at $\varphi^{(0)}_0 = 0$, $\varphi^{(0)}_q = 
\pi q/\lambda$ and $\varphi^{(0)}_{2q} = 2 \pi q/\lambda$ are always there, for any value of $\rho$. 
\item Compute the curvatures $\omega_n(\rho)$ of the potential $U(\varphi,\mu)$ at these zeros. 
\item Use the zeros previously determined to compute the classical masses $M_{ab}$ of the kinks connecting the various vacua
according to the topological formula (\ref{masskink}) 
\EQ
M_{ab} \,=\,\mid W(\varphi_a^{(0)}) - W(\varphi_b^{(0)}) \mid \,\,\,.
\EN 
The quantum corrections may change these values but in the semi-classical approximation ($\lambda \rightarrow 0$) one 
shall expect these corrections to be small. 
\item 
For each vacuum $\mid {\bf a}\rangle$, identify the kink $\varphi_{ab}$ with the {\em lowest} mass $M^*_{ab}$. 
\item Use the two previous data to compute for each vacuum   $\mid {\bf a} \rangle$ the quantity $\xi_a \,=\,\omega_a/(\pi M_a^*)$. 
\item The number of stable bosonic and fermionic particles at each vacuum $\mid {\bf a}\rangle$ is then given by 
$N_a \,=\left[\frac{1}{\xi_a}\right]$ and their mass is obtained by using the semi-classical formula (\ref{universalmassformula}) or (\ref{spectrofermion}). 
\item Identify the critical values $\rho_m^{(c)}$ when two zeros collide and disappear afterward. At these values the related kinks become massless and this phenomenon produces a singularity in certain $\xi_a$'s, as discussed in more detail in the next Section. 
\end{enumerate} 
It is interesting to see how this protocol determine the evolution of the spectrum in two simple but significant examples.
 
\section{Vacua Realization of SUSY} 
\label{vacuaSUSY}
\noindent
The disappearence of the vacua moving the parameter $\rho$ poses the question how SUSY is realised on the various vacua and 
whether it can get spontaneously broken in some of them even though it was originally exact on all the initial vacua. The answer is generically  
affermative and this because, once two zeros collide at the position $\varphi_c^{(0)}$ at some $\rho_c$, for $\rho = \rho_c +\epsilon$ 
we have $V(\varphi_c^{(0)},\rho_c+\epsilon) > 0$. This means that SUSY will be spontaneously broken at this point of the field space and, in the presence of other zeros, the new minimum at $\varphi_c^{(0)}$ becomes a meta-stable vacuum state. 
There is a notable exception to this scenario and it is related to the arithmetic properties of the ratio $\omega$ \citep{GMSUSYKINK}. Namely, when $(q-p)$ is an odd number, among the pairs of colliding zeros there will always be a couple of them placed just on the right and on the left of 
$\varphi^{(0)}_q = \pi q/\lambda$: when these zeros collide at some critical value $\rho_c$ they strangle the zero at $\pi q/\lambda$ that is in between, so that $\varphi^{(0)}_q$ remains a zero even after their collision.  This situation does not happen instead when $(q-p)$ is an even number. 

Since in the ultraviolet, all multiple SUSY Sine-Gordon models may be regarded as perturbation of the Superconformal Field Theory with central charge 
$c = 3/2$, it was argued in \citep{GMSUSYKINK}) that these theories give rise in general to a sequence of phase transitions that locally recalls the spontaneously symmetry breaking that occurs in the Tricritical Ising Model: moreover, when $(q-p)$ is an odd number, there is also an extra phase transition similar to the one of the gaussian model.    
\begin{figure}[t]
\psfig{figure=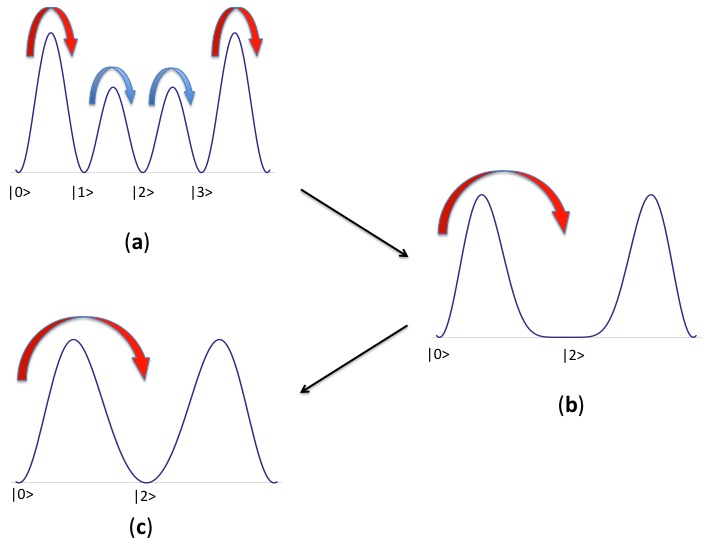,height=10cm,width=15cm}
\caption{{\em Evolution of the potential $V(\varphi,\rho)$ for the SUSY Double Sine-Gordon model: (a) $\rho = 0.3$; (b) $\rho = 0.5$ and 
$\rho=0.8$. In the figures there are drawn for simplicity only the kinks, in red the long-kinks, in blue the short-kinks}}
\label{DOUBLESINEGORDON}
\end{figure}

Let's now study in a certain detail two models which are somehow representative of the two classes where $(q-p)$ is an odd or an even number.
 
\subsection{Double Sine-Gordon Model}
\noindent
The first example is the Double SUSY Sine-Gordon model, with a bosonic potential given by 
\EQ
V(\varphi,\rho) \,=\,\frac{m^2}{2} \left(\frac{1-\rho}{\lambda} \sin\lambda\varphi + \frac{2 \rho}{\lambda} \sin\frac{\lambda \varphi}{2} \right)^2 
\,\,\,. 
\label{DoubleSSineGordon}
\EN 
In this case $\omega =\frac{1}{2}$ and $(q-p) = 1$. As we will see soon, this model has only one critical value of the parameter $\rho$ at 
$\rho_c \,=\,1/2$. Three different snapshot of the potential taken at $\rho < \rho_c$, $\rho = \rho_c$ and $\rho > \rho_c$, shown in Figure \ref{DOUBLESINEGORDON}, help in understanding and visualising the evolution of the model:
\begin{enumerate}[label=\alph*]
\item When $\rho < \rho_c$, in the enlarged interval of periodicity (\ref{newperiodicity}) there are $4$ stable vacua, identified as in Figure \ref{DOUBLESINEGORDON}.a. 
The vacua $\mid {0} \rangle$ and $\mid {1}\rangle$ are connected by long-kink/antikinks $\mid K_{10}^\pm\rangle$ and $\mid K_{10}^\pm\rangle$. 
The vacua $\mid 1 \rangle$ and $\mid 2 \rangle$, as well as $\mid 2 \rangle$ and $\mid 3 \rangle$, are connected instead by the short-kink/antikinks 
$\mid K_{12}^\pm\rangle$, $\mid K_{21}^\pm\rangle$ and $\mid K_{23}^\pm\rangle$, $\mid K_{32}^\pm\rangle$. In this model, the reflection symmetry with respect to the vacuum $\mid 2 \rangle$ provides an obvious identification of the latter four kinks. The vacua $\mid 0 \rangle$ and $\mid 2 \rangle$ correspond to the two fixed zeros $\varphi_0 = 0$ and $\varphi_2^{(0)} =\pi/\lambda$ of the potential (\ref{DoubleSSineGordon}) for all values of $\rho$. 
\item At $\rho =\rho_c =\frac{1}{2}$, the two vacua $\mid 1 \rangle$ and $\mid 3 \rangle$ collide, strangling between them the vacuum $\mid 2 \rangle$. Correspondingly the kinks $\mid K_{12}^\pm\rangle$ and $\mid K_{23}^\pm\rangle$ (together with their anti-kinks) become simultaneously massless. The critical value $\rho_c$ is identified as the value for  which the curvature at the vacuum $\mid 2 \rangle$ vanishes. The vacuum $\mid 2 \rangle$, 
according to our general analysis, survives the collision and the disappearance of the vacua $\mid 1 \rangle$ and $\mid 3 \rangle$.  
\item When $\rho > \rho_c$, there are only $\mid 0 \rangle$ and $\mid 2 \rangle$ as stable vacua in the theory, connected by a single kink 
$\mid K_{02}^\pm\rangle$ and the corresponding antikink  $\mid K_{20}^\pm\rangle$. 
\end{enumerate} 
Let's see how the spectrum at each vacuum evolves by varying $\rho$, choosing a particular value of $\lambda$, say  
$\lambda \,=\, 2 \sqrt{\pi/11} $. It is worth saying that the choice of any other value of $\lambda$ does not change the overall picture but just its 
details, as will become soon clear from the discussion below. 

\begin{figure}[t]
\psfig{figure=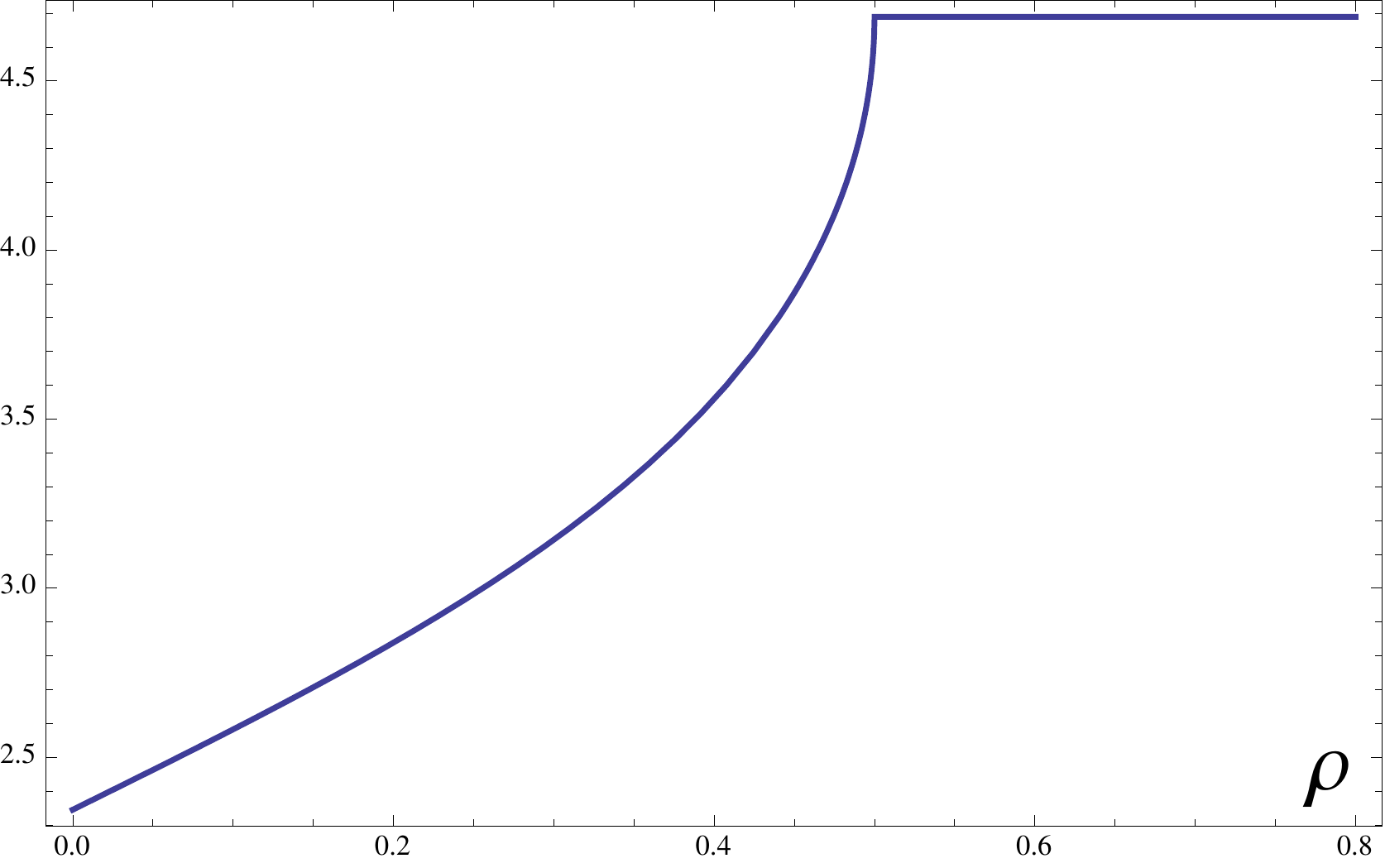,height=3.5cm,width=6cm}
\caption{{\em Root relative to the vacuum $\mid 1 \rangle$ of the SUSY Double Sine-Gordon as a function of $\rho$. For $\rho > \rho_c = \frac{1}{2}$, 
this root coincides with $\varphi_2^{(0)}$. }}
\label{nontrivialzero}
\end{figure}

We need first to determine the position $\varphi_1^{(0)}$ of the vacuum $\mid 1\rangle$ (and by symmetry with respect to the vacuum $\mid 2\rangle$, the other zero $\varphi_3^{(0)} \rangle$). This can be done numerically, finding the non-trivial zero of the potential (\ref{DoubleSSineGordon}): the result is plotted in Figure \ref{nontrivialzero}. Once $\varphi_1^{(0)}$ is known, using eq.\,(\ref{masskink}) we can compute the masses $M_{L,S}$ of the long and short kinks and their behavior versus $\rho$ is shown in Figure \ref{longshortkinkmass}. Computing as well as the curvature at each vacua and always using  the shortest kink for each vacuum, we can then determine the important parameters $\xi_a(\rho)$ and the numbers $N_a(\rho) = \left[\frac{1}{\xi_a}\right]$ of the bound states at each vacua. Let's now discuss what happens at each vacuum separately: 

\begin{figure}[b]
\psfig{figure=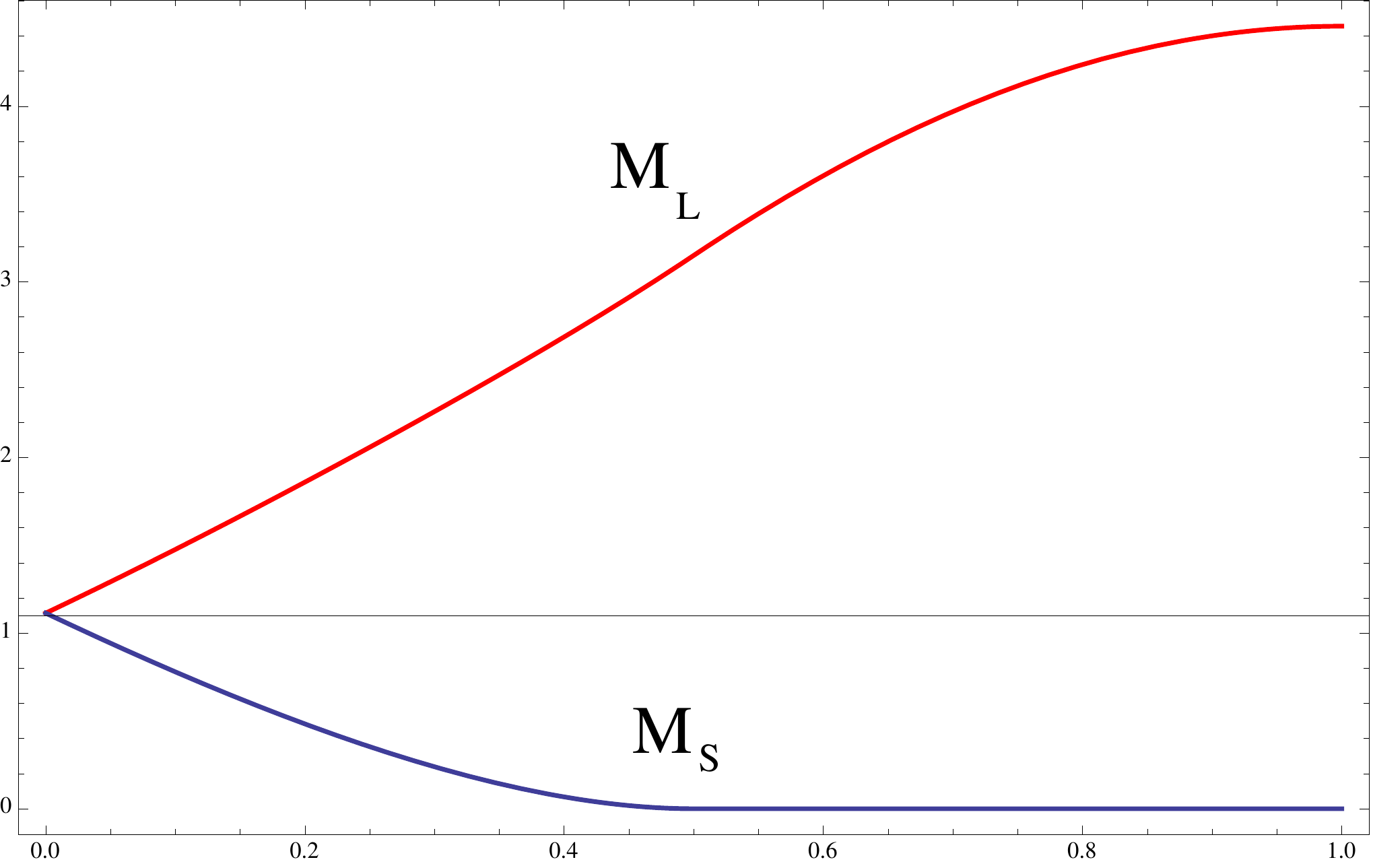,height=4cm,width=6cm}
\caption{{\em Mass $M_L$ (in red) and M$_S$ (in blue) of the long and short kink respectively, as function of $\rho$. Notice that 
$M_S = 0$ for $\rho > \rho_c$.}}
\label{longshortkinkmass}
\end{figure}

\begin{figure}[t]
\psfig{figure=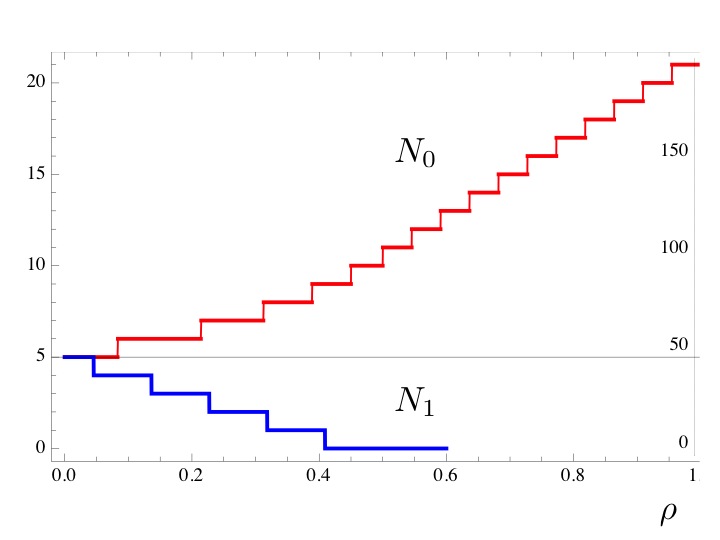,height=6cm,width=8cm}
\caption{{\em Number of bound states $N_0$ and $N_1$ as functions of $\rho$.}}
\label{scalettaa}
\end{figure}
\begin{figure}[b]
\psfig{figure=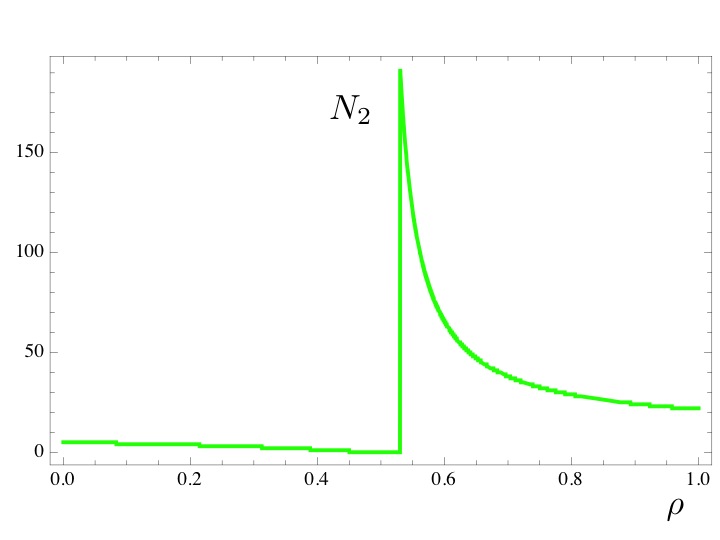,height=6cm,width=8cm}
\caption{{\em Number of bound states $N_2$ as function of $\rho$. Notice that 
$N_2$ diverges at $\rho = \rho_c$.}}
\label{scalettab}
\end{figure}
\begin{figure}[t]
\psfig{figure=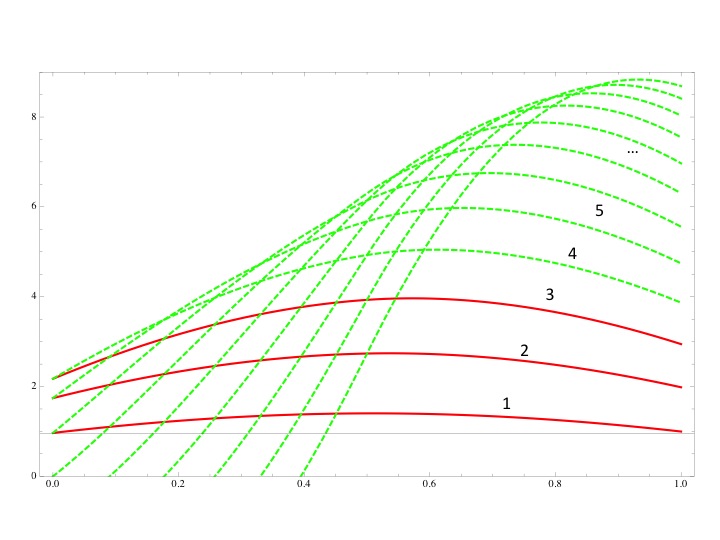,height=6cm,width=10cm}
\caption{{\em Evolution of the various masses of the bound states $N_0$ at the vacuum in the origin versus $\rho$. In solid red line those which are 
always present in the flow, where in green dashed lines those corresponding to the additional bound states which start to appear only once $\rho$ overpasses the special values $\rho^*_m$, even though in the Figure are plotted for all values of $\rho$.}}  
\label{boundstateorigin}
\end{figure}
\begin{enumerate}
\item the vacuum $\mid 0 \rangle$ exists for all values of $\rho$, therefore it makes sense to consider $N_0(\rho)$ along the entire flow. As shown in Figure \ref{scalettaa}, in this case $N_0(\rho)$ is a increasing staircase function which jumps discontinuously each time that $\rho$ overpasses certain special values $\rho^*_m$. At the beginning of the flow, the number of bound states is fixed by the first frequency $\lambda$ and it is given by $N_0(0) =3$, while at each $\rho^*_m$ a new bound state enters the spectrum coming from the continuum, so that at the end of the flow the total number of bound states is $N_0(1) = 13$, which is the number of bound state of the SUSY Sine-Gordon model relative to the second frequency $\frac{\lambda}{2}$. 
\item the vacua $\mid 1 \rangle$ and $\mid 3 \rangle$ exist only for $\rho < \rho_c$ and therefore it makes sense to talk about the number of the bound states $N_1(\rho)$ on top of these vacua only for this range of $\rho$. For $\rho > \rho_c$ we may regard this number to be instead equal to zero. The result is the step-wise decreasing function also shown in Figure \ref{scalettaa}, which starts as before from $N_1(0)=3$ and ends down then to zero.   
\item the number of bound states $N_2(\rho)$ on the vacua $\mid 2 \rangle$ presents a discontinuity at $\rho = \rho_c$ for the merging of three vacua at this value of $\rho$. For $\rho < \rho_c$, this number $N_2(\rho)$ starts at $\rho=0$ with the value 3 and decreases toward zero at $\rho_c$, similarly to $N_1(\rho)$.  But at $\rho_c$ this number diverges, simply because: (a) the curvature of this vacuum at that value vanishes and (b) the mass of the kink that has to be employed to compute $\xi_a$ for $\rho > \rho_c$ is no longer $M_S$ (that is zero) but $M_L$, which is instead finite. So, for $\rho > \rho_c$, $N_2(\rho)$ decreases from values infinitely large toward the final value $N_2(1) = 13$ dictated by the second frequency of the model, see Figura \ref{scalettab}. 
\end{enumerate}

With all these information one can then apply the semiclassical mass formulas to compute the actual masses of the bound states at each vacuum. 
Here we present the result only for the vacuum in the origin, Figure \ref{boundstateorigin}. Notice that the three lowest lines are present all over the flow while there is a cascade of additional bound states coming from the continuum each time that $\rho$ overpasses the values $\rho^*_m$. 

\begin{figure}[b]
\psfig{figure=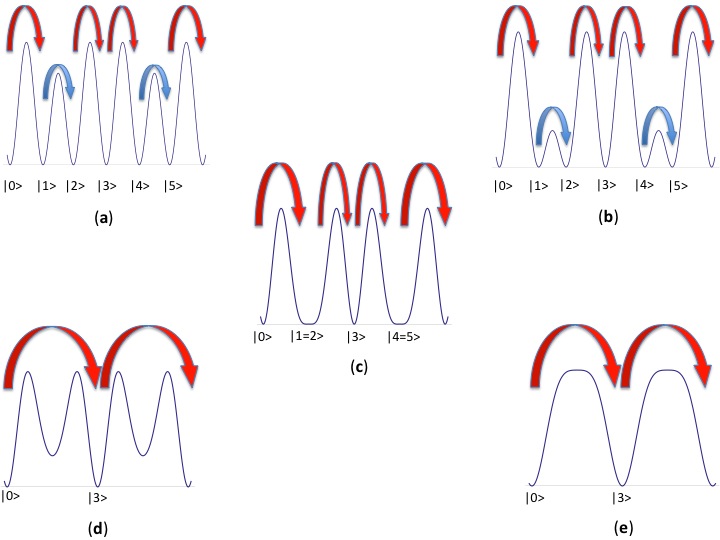,height=9cm,width=14cm}
\caption{{\em Evolution of the potential $V(\varphi,\rho)$ for the SUSY Triple Sine-Gordon model: (a) $\rho = 0.02$; (b) $\rho = 0.05$; 
(c) $\rho_c =0.2$; (d)$ \rho=0.4$ and $\rho=0.8$. In the figures there are drawn for simplicity only the kinks, in red the long-kinks, in blue the short-kinks}}
\label{TripleSGPotential}
\end{figure}

\subsection{Triple SUSY Sine-Gordon Model}

\begin{figure}[t]
\psfig{figure=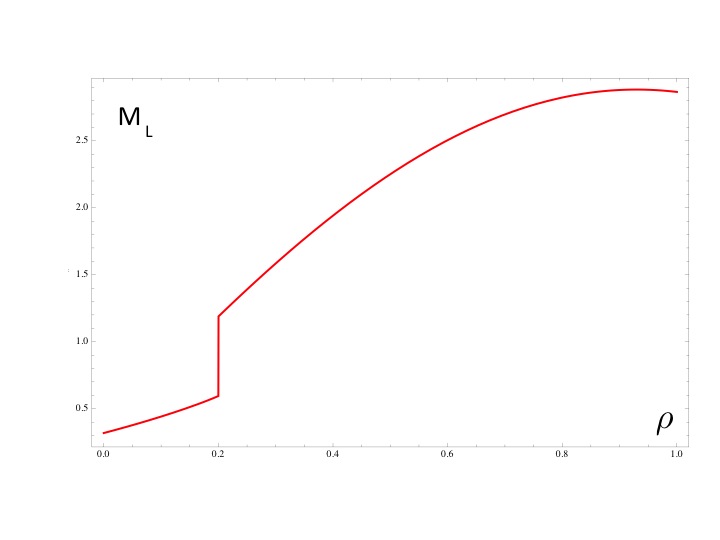,height=6cm,width=8.5cm}

\caption{{\em Mass $M_L$ of the long kink connecting the vacuum $\mid 0 \rangle$ to its nearest one. 
respectively, as function of $\rho$. Notice the discontinuity of the function at $\rho_c$.}}
\label{longshortkinkmasstriple}
\end{figure}

\begin{figure}[b]
\psfig{figure=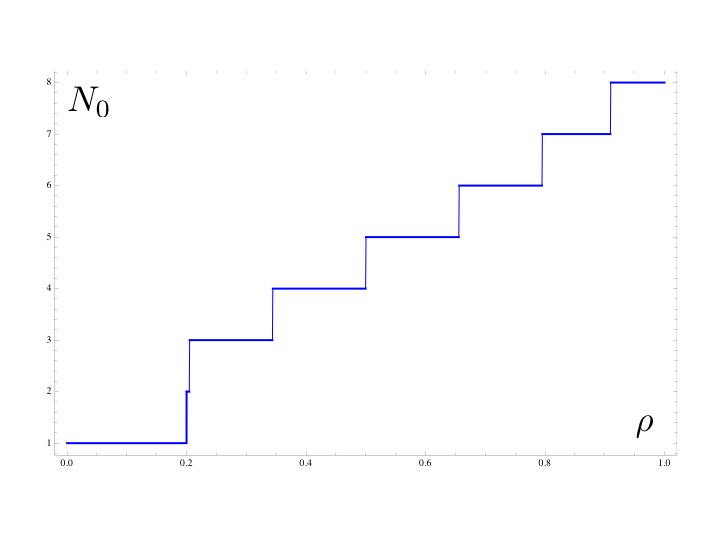,height=6cm,width=8cm}
\caption{{\em Number of bound states at $\mid 0 \rangle$ and $\mid 3 \rangle$. Notice the jump of 2 units at $\rho_c$.}}
\label{jump2}
\end{figure}

This is the model (\ref{newpotential}) where the ratio of the frequency is $\omega = 1/3$. The landscape of the potential of this model for various values of $\rho$ can be found in Figure \ref{TripleSGPotential}. One can see that, contrary to the previous model, here when two vacua collide at $\rho_c$ ($\mid {\bf 1} \rangle$ and $\mid {\bf 2} \rangle$, as well as $\mid {\bf 4} \rangle$ and $\mid {\bf 5} \rangle $), immediately after the potential is lift up, alias the two vacua disappear and nothing remains behind. This has the striking consequence that, not only the short kinks disappear at the critical point $\rho_c$ but also that the long kink immediately after the transition is made of the {\rm two previous long kinks}. This creates a discontinuity in the value of the mass of the lowest kink that has to be employed in the computation of $N_0(\rho)$ and $N_3(\rho)$, as can be seen from Figure \ref{longshortkinkmasstriple}. Correspondingly, there is a jump of 2 units in the number of bound states that piled up on the vacua $\mid {\bf 0} \rangle$ and $\mid {\bf 3} \rangle$, see Figure \ref{jump2}. So, the breaking of SUSY at the colliding vacua $\mid {\bf 1} \rangle $ and $\mid {\bf 2} \rangle$ (and the companion $\mid {\bf 4} \rangle$ and $\mid {\bf 5} \rangle $) has an effect on the spectrum of the nearby vacua where SUSY is instead still exact.

\section{Conclusions}
\label{conclusions}
\noindent
In this paper we have derived the semi-classical formula (\ref{generalexpressionmassfermion}) for the fermionic spectrum of an interacting QFT of boson and fermion. Together with the formula (\ref{universalmassformula}) for the bosons, these expressions provide interesting information on those quantum field theories made of degenerate vacua. Notice that the formula (\ref{generalexpressionmassfermion}) remains valid also when the fermion is of Dirac type rather than Majorana, modifying correspondingly the expansion of the fermion in the presence of the kink, eq.\,(\ref{fermionexpansion}), 
and using the more general relation between matrix elements given in eq.\,(\ref{matricDiracfermion}). So, these semi-classical formulas allow us to identify the particle excitations and estimate one of their most important characteristics, i.e. their mass. Of course, in absence of integrability, not all the particles entering the semi-classical formulas are stable: to determine which of them will decay simply involves to identify the proper threshold of the decay channel -- a task not particularly difficult but which we did not pursue here. The main difference with respect to the bosonic case analysed in \citep{GMneutral} is the presence of the shift $\eta_a$ in the fermionic formula  (\ref{generalexpressionmassfermion}) with respect to the bosonic one (\ref{universalmassformula}), which may influence the value of the thresholds. However it is natural to think that, as in the bosonic case \citep{GMneutral}, only few excitations will be generally stable on each vacua, all the others become resonances. 

With the fast advances of experimental works in quantum world -- some of the recent achievements were briefly underlined in the Introduction -- there may be soon the appealing perspective to realise many of the quantum field theories on a desk of atomic or material science laboratory. There has been in literature various proposals to address Majorana fermions in a context similar to the one discussed here, see for instance \citep{Stern} for 
a proposal of observing Majorana bound states of Josephson vortices in topological superconductors, or \citep{Affleck} for the possibility to observe  
an emergent Supersymmetry from strongly interacting Majorana zero modes in the Tricritical Ising Model. If and when these or other proposals could be implemented experimentally , it would be extremely intriguing to check some of the features discussed here, among which: the patter of the bound states on the various vacua in a situation of local breaking of supersymmetry in models as the multiple Sine-Gordon theories; or, in a situation of asymmetric potential, the swapping of the bosonic and fermions excitations present on the two vacua.

\vspace{1cm}
\begin{flushleft}\large
\textbf{Acknowledgements}
\end{flushleft}

I am very grateful to Michele Burrello for interesting discussions and for his careful reading of the manuscript. 
This work acknowledges the IRSES grant FP7-PEOPLE-2011-IRSES QICFT 295234.

\end{document}